\documentclass[prb,twocolumn,superscriptaddress]{revtex4}% Physical Review B
\usepackage[utf8]{inputenc}
\usepackage{graphicx}% Include figure files
\usepackage{dcolumn}% Align table columns on decimal point
\usepackage{bm}% bold math
\usepackage{color}
\usepackage{epstopdf}
\usepackage{subfigure}
\usepackage{times}

\begin{document}

\title{Optical and structural study of the pressure-induced phase transition of CdWO$_4$}

\author{J. Ruiz-Fuertes}
\email{javier.ruiz-fuertes@uv.es}
\affiliation{MALTA-Consolider Team. Departament de F\'{i}sica Aplicada-ICMUV, Universitat de Val\`{e}ncia, Dr. Moliner 50, 46100 Burjassot, Valencia, Spain}
\author{A. Friedrich}
\affiliation{Institut f\"ur Geowissenschaften, Goethe-Universit\"at Frankfurt, Altenh\"oferallee 1, 60438 Frankfurt am Main, Germany}
\affiliation {Institut f\"ur Anorganische Chemie, Julius-Maximilians-Universit\"at W\"urzburg, Am Hubland, 97074 W\"urzburg, Germany}
\author{D. Errandonea}
\author{A. Segura}
\affiliation{MALTA-Consolider Team. Departament de F\'{i}sica Aplicada-ICMUV, Universitat de Val\`{e}ncia, Dr. Moliner 50, 46100 Burjassot, Valencia, Spain}
\author{W. Morgenroth}
\affiliation{Institut f\"ur Geowissenschaften, Goethe-Universit\"at Frankfurt, Altenh\"oferallee 1, 60438 Frankfurt am Main, Germany}
\author{P. Rodr\'{i}guez-Hern\'{a}ndez}
 \affiliation {Instituto de Materiales y Nanotecnolog\'{i}a, Departamento de F\'{i}sica, Universidad de La Laguna, La Laguna, 38205 Tenerife, Spain}
\author{A. Mu\~{n}oz}
 \affiliation {Instituto de Materiales y Nanotecnolog\'{i}a, Departamento de F\'{i}sica, Universidad de La Laguna, La Laguna, 38205 Tenerife, Spain}
 \author{Y. Meng}
 \affiliation {HPCAT, Carnegie Institution of Washington, Bldg. 434E, 9700 S. Cass Avenue, Argonne, IL 60439, USA}
\date{\today}% It is always \today, today,

\begin{abstract}
The optical absorption of CdWO$_4$ is reported at high pressures up to
23 GPa. The onset of a phase transition was detected at 19.5 GPa, in
good agreement with a previous Raman spectroscopy study. The crystal
structure of the high-pressure phase of CdWO$_4$ was solved at 22 GPa
employing single-crystal synchrotron x-ray diffraction. The symmetry
changes from space group $P$2/$c$ in the low-pressure wolframite phase
to $P2_1/c$ in the high-pressure post-wolframite phase accompanied by
a doubling of the unit-cell volume. The octahedral oxygen coordination
of the tungsten and cadmium ions is increased to [7]-fold and
[6+1]-fold, respectively, at the phase transition. The compressibility of the low-pressure phase of CdWO$_4$ has been reevaluated with powder x-ray diffraction up to 15 GPa finding a bulk modulus of $B_0$ = 123 GPa. The direct band gap of the low-pressure phase increases with compression up to 16.9 GPa at 12 meV/GPa. At this point an indirect band gap
crosses the direct band gap and decreases at -2 meV/GPa up to 19.5 GPa where the phase transition starts. At the phase transition the band gap
collapses by 0.7 eV and another direct band gap decreases at -50
meV/GPa up to the maximum measured pressure. The structural stability
of the post-wolframite structure is confirmed by \textit{ab initio}
calculations finding the post-wolframite-type phase to be more stable
than the wolframite at 18 GPa. Lattice dynamic calculations based on space group $P2_1/c$ explain well the Raman-active
modes previously measured in the high-pressure post-wolframite
phase. The pressure-induced band gap crossing in the wolframite phase
as well as the pressure dependence of the direct
band gap in the high-pressure phase are further discussed with respect to the calculations.

\end{abstract}

\maketitle

\section{INTRODUCTION}
Nowadays used as a scintillating detector in x-ray tomography
\citep{rathe06}, high-energy particle physics \citep{mikha10}, and
dosimetry devices \citep{silva12}, the wide-band gap (4 eV)
semiconductor cadmium tungstate (CdWO$_4$) has been extensively
studied during the last three decades. It possesses a high light yield
emission when hit by $\gamma$ particles or x-rays and despite its long
scintillation time (12-15 $\mu$s) \citep{burac96}, it played a key role in
the discovery \citep{danev03} of the natural alpha activity in
$^{180}$W. Also, the long decay time of the radiation created by
self-trapped Frenkel excitons of CdWO$_4$ makes this material a test
bench for studying exciton-exciton interactions in semiconductors
\citep{kirmm09}. In order to improve the versatility of
CdWO$_4$ as a scintillating material, understanding how doping
\citep{novos12} or externally modifying its interatomic distances
affect its electronic structure, are of interest. In this context,
pressure is an efficient tool to correlate changes in the bond
distances with electronic properties.

\begin{figure}
\centering
\includegraphics[height=95mm]{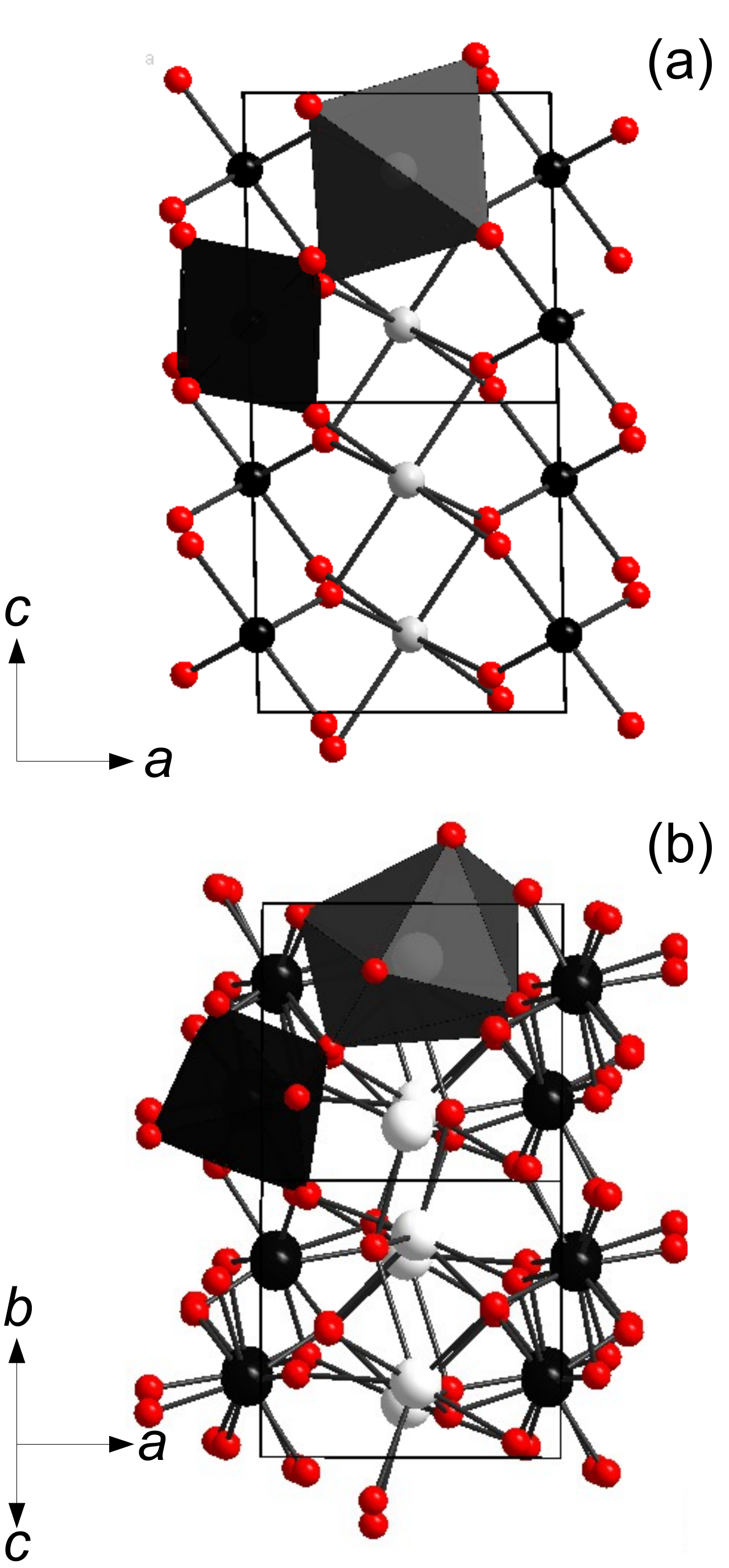}
\caption{\label{fig:fig1} Projections along (a) the [010] direction of
  the wolframite and (b) [011] direction of the post-wolframite-type
  structures of CdWO$_4$. The large white and black spheres represent
  the Cd and W atoms, respectively. The small spheres are the O
  atoms.}
\end{figure}

CdWO$_4$ crystallizes in a wolframite-type structure (space group $P$2/$c$)
at ambient conditions in which both Cd and W atoms are octahedrally
coordinated (Fig. \ref{fig:fig1}). Such a structure confers CdWO$_4$ a
direct band gap along the $Z$ direction of the Brillouin zone
\citep{abrah00,fujit08,ruizf12}. At high pressures, the conduction band,
mainly contributed by the 5$d$ W levels, moves up as a result of the
increase of repulsion, giving rise to a widening of the band gap at
12(1) meV/GPa up to at least 9.9 GPa \citep{ruizf12}. Previous
high-pressure polarized optical microscopic studies \citep{jayar95}
showed the emergence of aligned color domains at around 10 GPa in
CdWO$_4$, probably due to the low-hydrostatic conditions. However,
high-pressure Raman spectroscopy studies revealed that CdWO$_4$
remains in the wolframite-type structure up to at least 20 GPa
\citep{jayar95,lacom09}. Above this pressure the number of
Raman-active modes abruptly increases as a consequence of a phase
transition, interpreted by the coexistence of a triclinic
CuWO$_4$-type structure and a tetragonal scheelite-type structure,
according to the predictions \citep{lacom09}. The proposed
formation and coexistence of the two energetically similarly stable
high-pressure phases at the same pressure, though explained the number of the observed Raman modes, is controversial
since those Raman spectroscopy experiments were performed using a
single crystal \citep{lacom09}. In order to study such a phase
transition one should employ a structural technique on a single
crystal at the same experimental conditions. However, previous
single-crystal x-ray diffraction (SXRD) has been limited to 8.2 GPa
\citep{macav93}. Furthermore, previous electronic band structure
calculations have indicated a band crossing in CdWO$_4$ at 16 GPa, a pressure not
experimentally explored yet \citep{ruizf12} with optical absorption.

In this work we present a powder XRD and a SXRD study of CdWO$_4$ up to 15 GPa and 22 GPa, respectively, to
solve the structure of its high-pressure phase, and an optical absorption study up to 23
GPa to investigate the effect that the structural phase transition has
on the optical properties of CdWO$_4$. Finally, the stability of the
high-pressure phase of CdWO$_4$, the frequencies of its Raman modes,
and its electronic properties have been investigated using \textit{ab
  initio} calculations.

\section{EXPERIMENTAL DETAILS}
\label{exp}

For the non polarized optical absorption experiments at high pressure two
10-$\mu$m-thick samples of CdWO$_4$ were cleaved along the (010) plane
from a large single crystal obtained from the MTI Corporation. redIn this orientation the cross-polarization terms of the complex reflection are both zero \citep{jelli11}. The
samples were loaded in Neon pressure medium together with a ruby chip for pressure
calibration \citep{maohk78} in a membrane-type diamond anvil cell (DAC) with
500-$\mu$m culet anvils. The pressure chamber consisted of a hole
with a diameter of 200-$\mu$m drilled in an Inconel gasket previously preindented
to 45-$\mu$m thickness. The confocal optical setup used for the measurements consisted in a
deuterium-halogen lamp, fused silica lenses, two Cassegrain
objectives, and an UV-VIS spectrometer. The powder XRD experiments were carried out up to 15 GPa at the 16-IDB beamline of the HPCAT at the Advanced Photon Source (APS). The experiment was performed in a symmetric DAC with 480-$\mu$m diamond culets and a rhenium gasket with a hole of 150 $\mu$m was used as the pressure chamber. Neon was employed as pressure medium and the ruby fluorescence was used for pressure calibration. The monochromatic x-ray beam used had a wavelength of $\lambda$= 0.36783  Å and was  focused  down  to  $10 \times 10$ $\mu$m$^2$. For the SXRD experiments we
loaded a single-crystal sample together with a ruby chip and Neon was used as pressure medium in
Boehler-Almax DACs equipped with diamond culets of 350 $\mu$m diameter
and tungsten gaskets preindented to 45 $\mu$m thickness with a hole
of 160-$\mu$m diameter. The experiments were carried out at 15 and 22
GPa at the Extreme Conditions Beamline at PETRA III ($\lambda$ =
0.2925 {\AA} and 0.2904 {\AA}, respectively) using a PerkinElmer 1621
detector with a 8.3 $\times$ 9.6 $\mu$m$^{2}$ and 4 $\times$ 8
$\mu$m$^{2}$ beam and a sample to detector distance of 402.5 mm and
430.6 mm, respectively. The diffraction images were collected by
1$^{\circ}$ $\omega$-scanning. The image format was converted
according to the procedure described by \citet{rothk13} for further
processing with the CrysAlisPro software \citep{crysal} for indexing
reflections and intensity data reduction. The
crystal structure of the high-pressure phase of CdWO$_4$ exhibits a
pseudo-orthorhombic metric at 22 GPa. However, attempts to solve the
crystal structure in an orthorhombic space group were
unsuccessful. Finally, the crystal structure was successfully solved
in space group $Pc$ with SHELXS97-2 \citep{shel08} using direct
methods and was subsequently converted to space group $P2_1/c$ using
PLATON \citep{spek2009}. A pseudo-merohedral 2-component twin (55 \%)
related via a twin plane perpendicular to the $c$-axis with the twin
matrix (1 0 0, 0 1 0, 0 0 -1) was refined with SHELXL97-2
\citep{shel08}. The programs were used with the WinGX interface
\citep{farrugia1999}. The final refinements were carried out with
anisotropic displancement parameters for the tungsten and cadmium
atoms and isotropic ones for the oxygen atoms. The final residual
value R1 of the post-wolframite structure converged to 6.34 \% at a
data:parameter ratio of $\approx$ 26, which are excellent values for
single-crystal structure refinements of high-pressure data, and even
more for a twinned high-pressure phase. 

\section{COMPUTATIONAL DETAILS}

First principles calculations of the total-energy to study electronic structure and lattice dynamics were done within the framework of the density-functional theory (DFT) \citep{hohen64} and the pseudo-potential method using the Vienna \textit{ab initio} simulation package (VASP) \citep{kress93,kress94,kress96}. The exchange and correlation energy were used in the generalized gradient approximation (GGA) with the PBE functional \citep{perde96}. The projector-augmented wave (PAW) scheme \citep{bloch94,kress99} was adopted and the semicore $5p$ W electrons were also explicitly included in the calculations. The considered valence electron configuration was Cd 4$d^{10}$5$s^2$, O 2$s^2$2$p^4$, and W 5$d^4$6$s^2$. The set of plane waves employed extended up to the kinetic energy cutoff of 530 eV to deal with the O atoms in order to have highly accurate results. The Monkhorst-Pack \citep{monkh76} grid used for Brillouin-zone integrations ensured converged results to about 1 meV per formula unit. To perform the geometrical optimization we used 20 special $k$-points of the reciprocal space with symmetry $P2/c$, 14 $k$-points with $P2_1/c$, and 36 and 39 $k$-points for $P_1$ and $I4_1/a$ symmetries, respectively. In our simulations we obtain not only the energy and volume, we also obtain derivatives of the energy, forces, and stress. Hence, we also obtain the pressure from the \textit{ab initio} simulations. In fact when we relax for a volume a structure, the relaxed structure should have diagonal stress tensor (hydrostatic), and zero forces on the atoms. In the relaxed equilibrium configuration the forces on the atoms were lower than 0.002 eV/Å and the deviation of the stress tensor from a diagonal hydrostatic form was less than 0.1 GPa. Lattice dynamic calculations were carried out at $\Gamma$ using the direct forces constant approach. This method involves the construction of the dynamical matrix, that requires highly converged results on forces. Diagonalization of the dynamical matrix provides both the frequencies of the normal modes and their polarization vectors. This allows us to identify the irreducible representation and the character of the phonon modes at the zone center. More details of the calculations can be found in Refs. \citep{ruizf12,lacom09,ruizf10}

\section{EXPERIMENTAL RESULTS}
\subsection{Optical Absorption}

\begin{figure}
\centering
\includegraphics[width=0.35\textwidth]{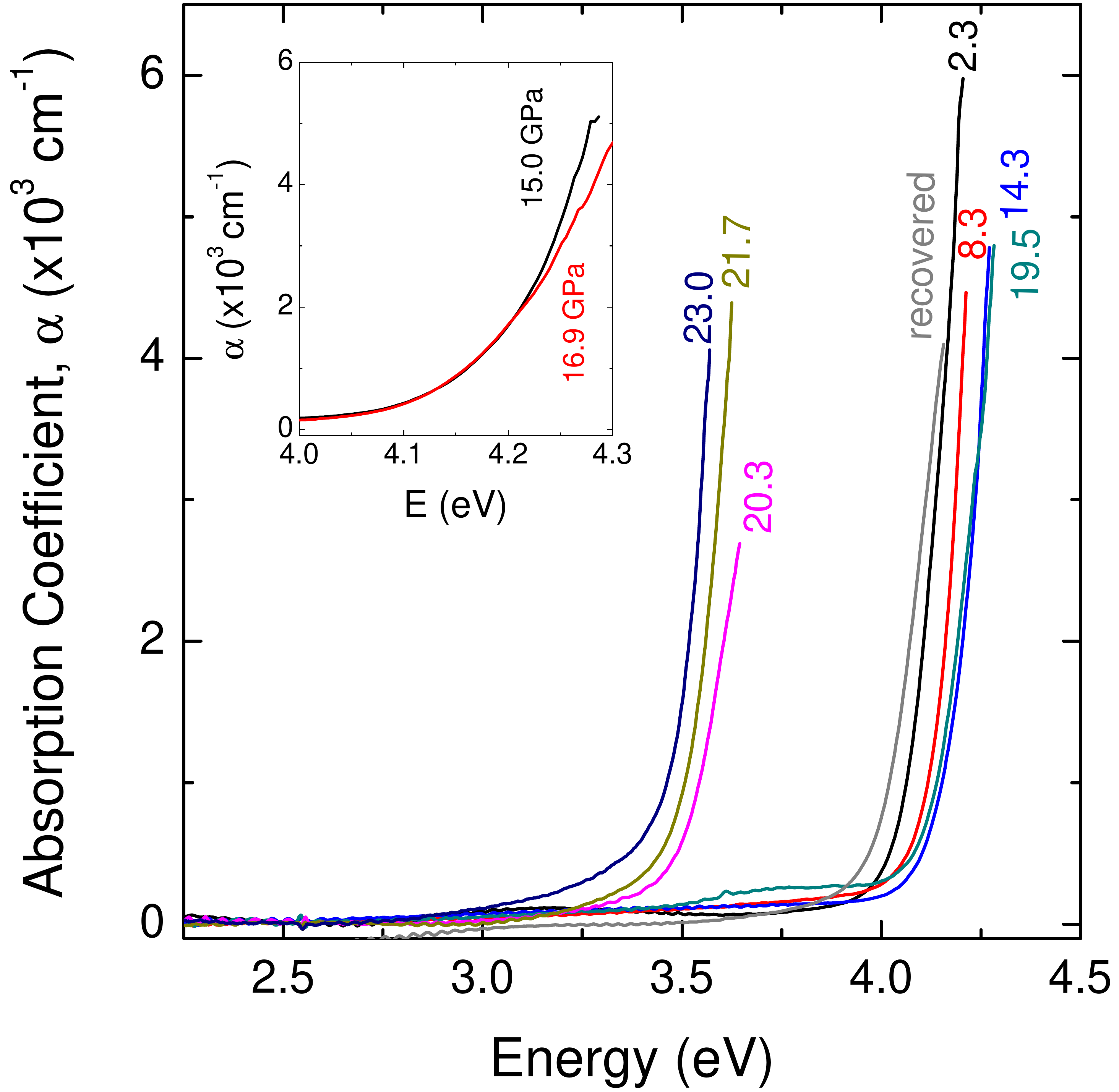}
\caption{\label{fig:fig2} Optical absorption spectra of CdWO$_4$ at different pressures up to 23 GPa. The inset shows the change of the shape of the absorption edge at 16.9 GPa compared to the spectrum at 15.0 GPa.}
\end{figure}

A selection of absorption-coefficient ($\alpha$) spectra of CdWO$_4$
at different pressures can be found in Fig. \ref{fig:fig2}. At ambient
pressure the absorption edge shows a steep increase up to a maximum
value of $\alpha \approx 6\times$10$^{3}$ cm$^{-1}$, in good agreement
with a direct band gap \citep{ruizf12}. As observed in previous
results up to 10 GPa \citep{ruizf12}, as pressure increases, the
absorption edge keeps the same shape and shifts from 4 eV to higher
energies up to 15 GPa. At 16.9 GPa the absorption edge reduces its
steepness and starts to downshift to lower energies with increasing pressure up to 19.5 GPa
when the sample exhibits a subtle color change and a low-absorbance
step-like absorption spectrum emerges at lower energy ($\approx$ 3.5
eV) in coexistence with the main absorption edge
(Fig. \ref{fig:fig2}). The shape of the newly emerged absorption spectrum indicates the contribution
of two distinct electronic transitions with different
absorbances. CdWO$_4$ undergoes a structural phase transition at 20
GPa according to Raman Spectroscopy \citep{lacom09}. Therefore, the
existence of an additional low-absorbance absorption edge at 19.5 GPa
can be attributed to the onset of the phase transition to the high-pressure phase, which at this pressure is a minor component.  This is confirmed at 20.3 GPa
with the loss of the high-energy absorption edge and an abrupt increase in the absorpbance of the low-energy absorption edge ($\alpha
\approx 4\times$10$^{3}$ cm$^{-1}$), indicating the end of the phase
transition. At higher pressures the absorption edge of the
high-pressure phase of CdWO$_4$ decreases in energy with further compression.

\begin{figure}
\centering
\includegraphics[width=0.35\textwidth]{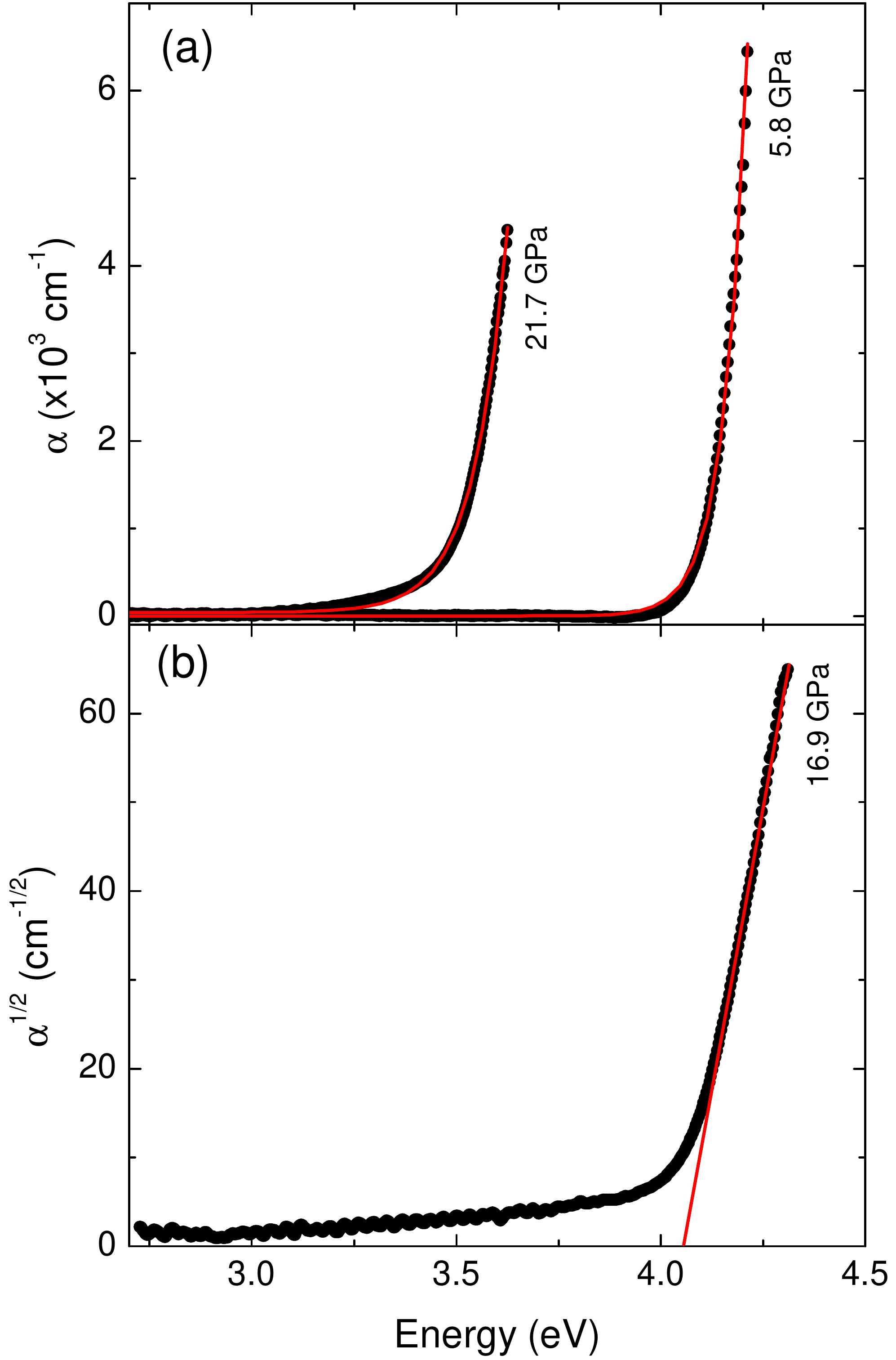}
\caption{\label{fig:fig3} (a) Fits to the Urbach's rule (red line) to two spectra at 5.8 and 21.7 GPa (black points) that correspond to the low-pressure and high-pressure phases of CdWO$_4$, respectively. (b) Linear dependence of $\alpha^{1/2}$ of the spectrum at 16.9 GPa showing the indirect nature of the band gap at this pressure.}
\end{figure}

The steep-shaped absorption edges observed in the low-pressure phase absorption edge up to 15
GPa and in the high-pressure phase above 20.3 GPa are typical of
direct band gaps. We have analyzed the absorption spectra in both
phases employing the direct-band gap Urbarch's rule $\alpha =
A_0\cdot\text{exp}[(h \nu - E_g)/ E_{\text{U}}]$, where $E_g$ is the
band gap, $E_{\text{U}}$ is Urbach's energy and is related to the
steepness of the absorption tail, and $A_0$ is an intrinsic constant
of the sample that accounts for the absorption of defects \citep{urbac53}. The fits to spectra of the low- and
high-pressure phases (Fig. \ref{fig:fig3}) confirm the direct nature
of the band gaps in both phases (except at 16.9 and 19.5
GPa). In hydrostatic conditions \citep{klotz09}, where uniaxial stresses do not deteriorate the sample, $E_{\text{U}}$ and $A_0$ can be assumed to remain constant under
pressure as long as the structure of the compounds and the density of
defects remain constant \citep{ruizf08}. In the case of CdWO$_4$ we
have obtained an $A_0$ value of 600 cm$^{-1}$ for both phases in good agreement with other wolframite-type compounds ($A_0
= 500$ cm$^{-1}$ for MgWO$_4$) \citep{ruizf12}. For the $E_{\text{U}}$, it takes
a value of 0.056 eV in the low-pressure phase and 0.082 eV in the
high-pressure phase. The higher value observed for $E_{\text{U}}$ in
the high-pressure phase indicates an increase of point defects usually
observed after a structural phase transition
\citep{erran06,lacom11,panch11}. Regarding the absorption edges at
16.9 GPa and 19.5 GPa, in Fig. \ref{fig:fig2} we can see that
  they lose steepness, redshift with pressure, and cross the absorption edges measured at lower
pressures. The linear dependence of $\alpha^{1/2}$ with
the photon energy (Fig. \ref{fig:fig3}) confirms the indirect nature
of the band gap at 16.9 and 19.5 GPa. After releasing pressure the recovered sample shows an absorption edge similar in shape and energy to the absorption edge of the low-pressure phase. This confirms the reversibility of the phase transition and the fact that pressure apparently does not induce defects in the used hydrostatic conditions. 

\begin{figure}
\centering
\includegraphics[width=0.35\textwidth]{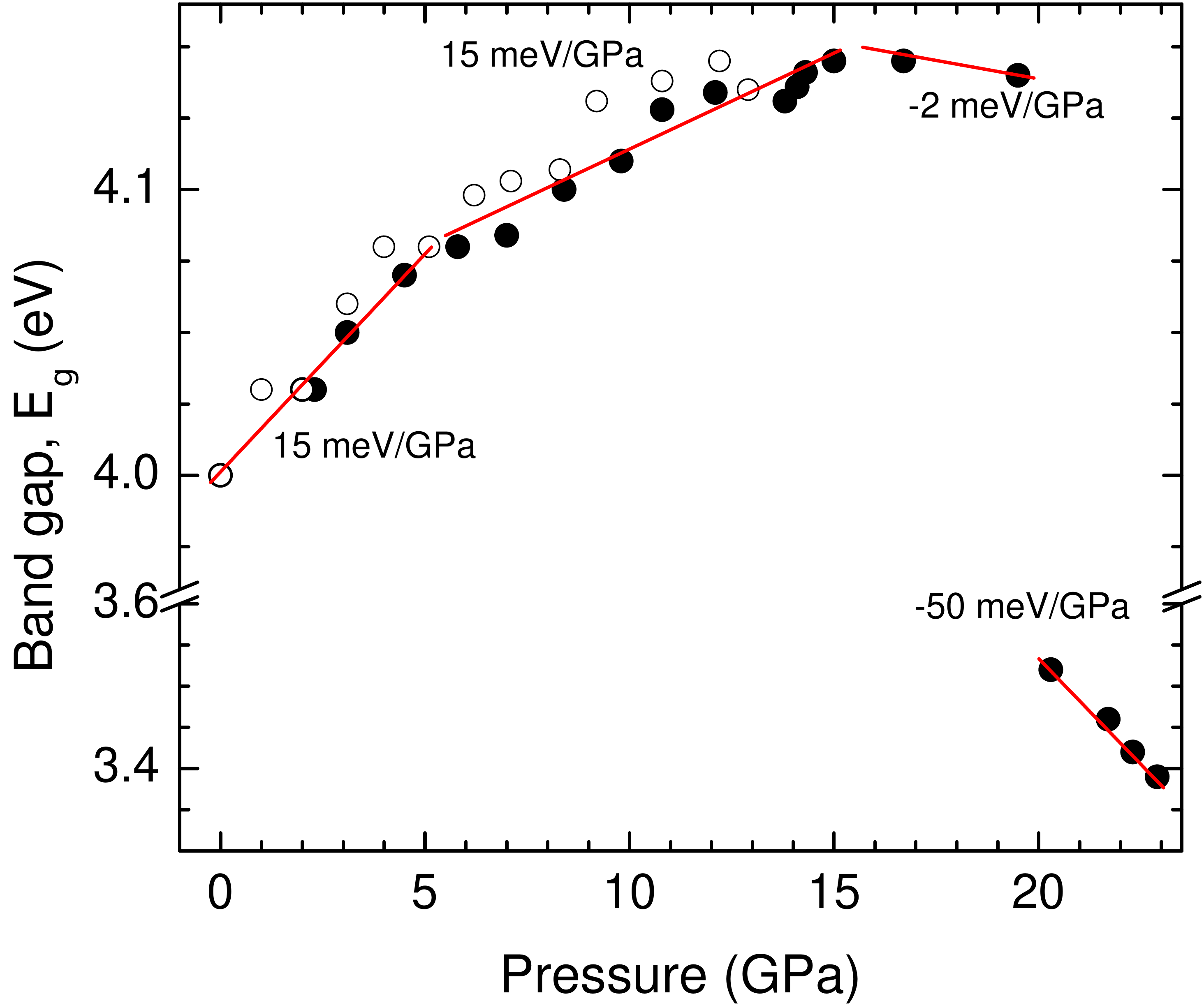}
\caption{\label{fig:fig4} Pressure dependence of the band gap energy of
  CdWO$_4$. Open circles represent data from Ref. \citep{ruizf12} while
  filled circles are from data collected in this study. The continuous
  red lines show the fits.}
\end{figure}

The values of the band gap energy obtained from the analysis of all measured spectra are shown in Fig. \ref{fig:fig4} together with the data from Ref. \citep{ruizf12}. We show that in the wolframite-type phase, the band gap increases up to 15 GPa  with a pressure coefficient d$E_g$/d$p$ of 12 meV/GPa in good agreement with previous data \citep{ruizf12}. Above 15 GPa, the change from a direct to an indirect band gap, associated to a band crossing previously reported \citep{ruizf12} and observed in other wolframites before the phase transition \citep{erran16}, implies a negative pressure coefficient of -2 meV/GPa. At the phase transition the band gap collapses by 0.7 eV, with the direct band gap of the high-pressure phase showing a negative pressure dependence of -50 meV/GPa.

\subsection{X-ray diffraction}

As explained in Sec. \ref{exp}, we carried out one powder XRD experiment \citep{support} up to 15 GPa and two SXRD experiments with CdWO$_4$. One of the two SXRD experiments was performed below and the other one above the phase transition at 15 GPa and 22 GPa, respectively.

The unit-cell parameters and atomic coordinates obtained from the structural refinements of the single crystals at those two pressures are given in table \ref{tab1}. \citep{support}

\begin{table}
\footnotesize
%\tiny
\renewcommand{\arraystretch}{1.5}
\caption{Lattice parameters, atomic coordinates, and equivalent or
  isotropic displacement parameters, respectively, of the experimental 
  and calculated wolframite-type and post-wolframite-type structures of CdWO$_4$. The lattice parameters are in Å, the $\beta$ angle is in
  degrees, the unit-cell volume is in Å$^3$, and the displacement parameters are in {\AA}$^2$.}
\centering { \tabcolsep=0.11cm
\begin{tabular}{cccccc}
\hline \hline
\multicolumn{2}{c}{} & \multicolumn{2}{c}{$P2/c$} & \multicolumn{2}{c}{$P2_1/c$} \\
\multicolumn{2}{c}{} & 15 GPa & 16 GPa & 22 GPa & 22.9 GPa \\
\multicolumn{2}{c}{} & \textbf{Exp.} & \textbf{Calc.} & \textbf{Exp.} & \textbf{Calc.} \\  
 \hline
  & $a$  & 4.9431(4) & 4.9795 & 5.1884(7)    & 5.2585 \\
  & $b$  & 5.498(5)  & 5.6553 & 6.1898(17)   & 6.2535 \\
  & $c$  & 4.9685(4) & 5.0110 & 7.5402(17)   & 7.66382 \\
  & $\beta$  & 93.136(6) & 92.26 & 90.01(2) & 90.67 \\
  & $V$  & 134.82(12) & 140.99 & 242.15(9) & 251.99 \\
  & $Z$  & 2 & 2 & 4 & 4\\
\hline
Cd& $x$ &  0.5 & 0.5   & 0.5162(3) &  0.51039\\
  & $y$ &  0.7126(1)   & 0.70415 & 0.7675(3) &  0.76775\\
  & $z$ &  0.25 & 0.25 & -0.0358(2) &   -0.02630\\
  & $U_{eq}$ & 0.007(2) & & 0.0117(3) & \\
W & $x$ &  0 & 0&  -0.04762(17)   &  -0.04827\\
  & $y$ &  0.1895(6) &0.18652 & 0.48387(13)    &  0.48930\\
  & $z$ &  0.25 & 0.25& 0.20900(13)    &  0.20798\\
  & $U_{eq}$ & 0.0047(15) & & 0.00936(19) & \\
O1& $x$ &  0.2112(15) & 0.20969 & 0.223(3)       &  0.22598\\
  & $y$ &  0.918(4) & 0.91069& 0.511(3)       &  0.50807\\
  & $z$ &  0.4645(14) & 0.45827& 0.034(2)       &  0.03741\\
  & $U_{iso}$ & 0.0079(11) & & 0.008(2) & \\
O2& $x$ &  0.2414(16) & 0.24264& -0.150(3)      &  -0.14766\\
  & $y$ &  0.397(5) & 0.38919& 0.790(3)       &  0.79670\\
  & $z$ &  0.3931(16) & 0.39126& 0.155(2)       &  0.14945\\
  & $U_{iso}$ & 0.0089(13)  & & 0.012(2) & \\
O3& $x$ &  & & 0.117(3)       &  0.11768\\
  & $y$ &  & & 0.670(3)       &  0.67751\\
  & $z$ &  & & 0.371(2)       &  0.36775\\
  & $U_{iso}$ & & & 0.010(2)  & \\
O4& $x$ &  & & -0.375(3)      &  -0.3123\\
  & $y$ &  & & 0.457(3)       &  0.46837\\
  & $z$ &  & & 0.300(3)       &  0.29774\\
  & $U_{iso}$ & & & 0.014(3) & \\
\hline \hline
\end{tabular}
}
\label{tab1}
\end{table}

\begin{figure}
\centering
\includegraphics[width=0.35\textwidth]{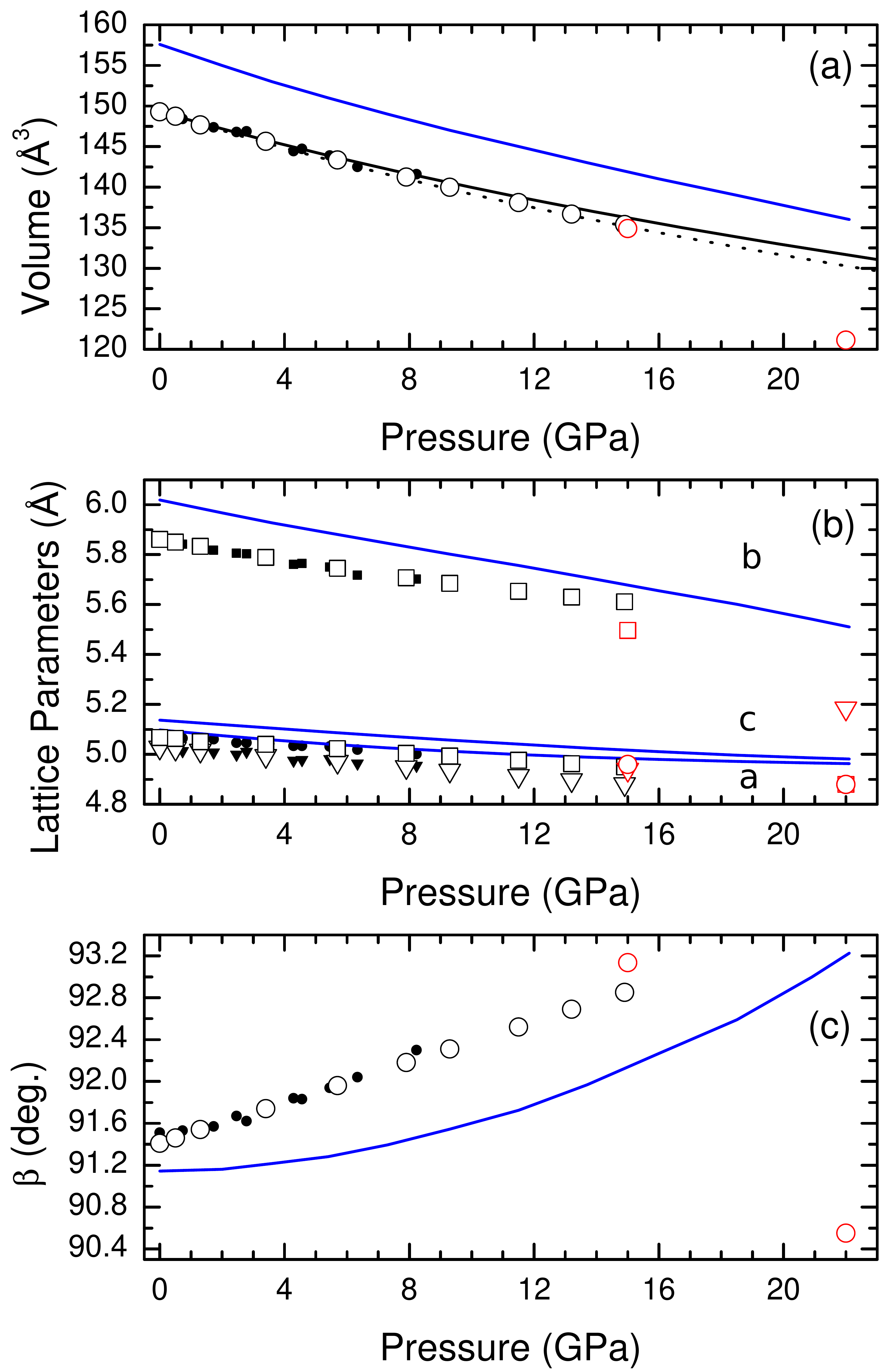}
\caption{\label{fig:fig5} Pressure dependencies of (a) unit-cell
  volume, (b) unit-cell axes, and (c) monoclinic $\beta$ angle of CdWO$_4$. Data shown by solid symbols are from \citet{macav93} (black) and data shown by black (red) open symbols are from the powder XRD (SXRD) experiment. Fits of the data by \citet{macav93} and of the
  data including our powder diffraction data up to 15 GPa to Birch-Murnaghan
  equations of state (EOS) of second order are drawn by black solid and
  dashed lines, respectively. At the highest
  pressure of 22 GPa, only half of the unit-cell volume of the
  post-wolframite phase is shown to compare with that of the
  wolframite phase. Lattice parameters at 22 GPa were
  transformed to those of the respective wolframite cell parameters
  for comparison. Blue lines represent the calculations.}
\end{figure}

In Fig. \ref{fig:fig5} we show the pressure dependence of the unit-cell volume and lattice parameters from a previous study by \citet{macav93} and from our powder XRD and SXRD experiments. Previous SXRD data \citep{macav93} had been limited to 8.2 GPa. We have increased the pressure range to 15 GPa for the low-pressure phase and observed that the bulk modulus of this phase \citep{macav93} had been overestimated (Fig. \ref{fig:fig5}a). Using a second order Birch-Murnaghan equation of state to fit the combined data of \citet{macav93} and ours, we obtain a bulk modulus $B_0 = 123$ GPa instead of 136 GPa that had previously been reported by \citet{macav93}. Regarding the lattice parameters (Fig. \ref{fig:fig5}b) we confirm the observation by \citet{macav93} that the $b$ axis is more compressible than the other axes. Similarly to MnWO$_4$ \citep{ruizf15}, the $\beta$ monoclinic angle increases with pressure up to the phase transition.

The crystal structure of the high-pressure post-wolframite phase of
CdWO$_4$ was solved at 22 GPa from the SXRD data. While the $a$-axis is increased with
respect to the low-pressure structure at 15 GPa, the \textbf{b} and \textbf{c} basis
vectors of the post-wolframite cell relate to the diagonal directions
[0$\bar{1}$1] and [0$\bar{1}\bar{1}$] in the \textbf{bc} plane of the
wolframite cell, respectively. This is also reflected in a doubling of
the unit cell volume of the post-wolframite phase with respect to that
of the wolframite phase, and hence of the formula units from 2 to
4. The transformation can be described by the transformation matrix [1
  0 0, 0 -1 1, 0 -1 -1]. The space group symmetry changes from
$P$2/$c$ in the wolframite phase to $P$2$_1$/$c$ in the
post-wolframite phase ($\vec{b'}$ = -$\vec{b}$+$\vec{c}$,
$\vec{c'}$=-$\vec{b}$-$\vec{c}$) at the phase transition accompanied
by a change in the monoclinic $b$-axis direction. The
orientation relation between the wolframite and the post-wolframite
phase was confirmed by a comparison of the orientation matrices of the
same crystal at pressures below and above the phase transition. There
is no group-subgroup relationship between the wolframite ($P2/c$) and
post-wolframite ($P2_1/c$) space groups. A comparison of half of the
post-wolframite unit-cell volume at 22 GPa with the wolframite unit-cell volume extrapolated from the equation of state to 22 GPa
indicates a volume reduction of about 8 \% at the phase transition
(Fig. \ref{fig:fig5}a). Such a huge volume drop points towards a
first-order type of phase transition. Most of the volume compression
is achieved by the compression of the wolframite $b$ axis, which also
compensates for the significant increase of the $a$ axis
(Fig. \ref{fig:fig5}b).  The crystal structure is severely reorganized
at the phase transition, which is also expressed in a strong
broadening of the single crystal reflections in the high-pressure phase. At the phase
transition, the oxygen coordinations of cadmium and tungsten cations
are increased from [6] to [6+1] and [7], respectively
(Fig. \ref{fig:fig1}).

%distortion of the structure - lattice parameters! 

Our crystal structure solution of the post-wolframite phase of
CdWO$_4$ from single-crystal x-ray diffraction clearly contradicts the
earlier assumption of the coexistence of two high-pressure phases by
\citet{lacom09}. In that work the formation of both a high-pressure
phase with a CuWO$_4$-type structure (S. G. $P\bar{1}$) and another
one of the scheelite-type (S. G. $I4_1/a$) had been proposed using
\textit{ab initio} calculations in order to interpret the
experimentally observed Raman modes \citep{lacom09}. Using our
experimentally determined crystal structure, the Raman-mode assignment
is revisited in section \ref{section:Raman} and the calculation of the
electronic band structure is presented in section
\ref{section:bandgap}, where it is discussed with respect to the
pressure-induced behavior of the band gap across the
phase transition.

\section{THEORETICAL RESULTS}

\subsection{Phase Stability}
\begin{figure}
\centering
\includegraphics[width=0.40\textwidth]{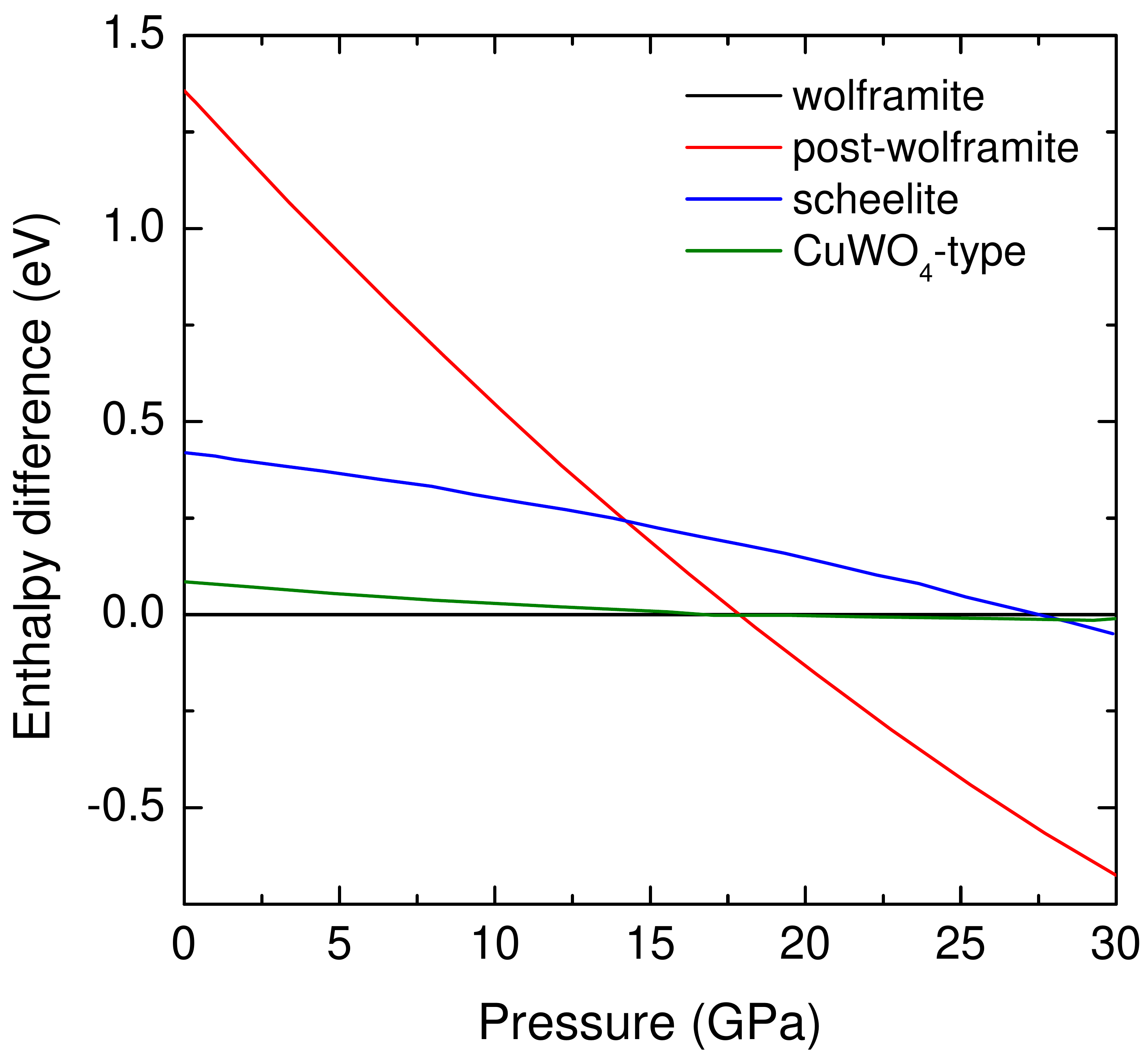}
\caption{\label{fig:fig6} Calculated enthalpy difference of CuWO$_4$-,
  scheelite-, and post-wolframite-type phases of CdWO$_4$ with
  respect to the enthalphy of the low-pressure wolframite-type
  one. The enthalpy of wolframite is taken as reference}.
\end{figure}

The calculated enthalpy differences of CdWO$_4$ for the scheelite-type
(S. G. $I4_1/a$), CuWO$_4$-type (S. G. $P\bar{1}$), and
post-wolframite phase (S. G. $P2_1/c$) as solved in this study are
plotted with respect to the enthalpy of the wolframite-type structure
(S. G. $P2/c$) in Fig. \ref{fig:fig6}.

At high pressures, \citet{lacom09} reported from their calculations that
the scheelite-type structure would become slightly more stable than
the wolframite above 21.2 GPa. This is a pressure slightly higher than that of
the phase transition onset (19.5 GPa) observed in the optical absorption study, but can be considered as a reasonable agreement. However,
our calculations show that the post-wolframite structure of CdWO$_4$,
solved with SXRD at 22 GPa in this work, becomes more stable at 18
GPa than both the
wolframite-type structure and the CuWO$_4$-type structure, supporting our experimental structural solution.

In order to obtain the bulk moduli of both low- and high-pressure phases from the calculated data, we have also employed second order Birch-Murnaghan fits for a better comparison with the experimental data. The bulk modulus of the wolframite phase of CdWO$_4$ from the calculations is $B_0=111$. This is 9\% lower than our updated value using combined experimental data from our study up to 15 GPa and from \citet{macav93}, and 18\% lower than the $B_0$ obtained by \citet{macav93} up to 8.2 GPa. The bulk modulus obtained in our calculations for the post-wolframite phase it is $B_0=114.5$ GPa.   

\subsection{Raman modes}\label{section:Raman}

Since the previous structural prediction of the post-wolframite phase of CdWO$_4$ had been based on the assignment of experimentally observed Raman modes, we have calculated frequencies and pressure coefficients ($\partial \omega/\partial P$) of the Raman
active modes of CdWO$_4$ at 26.1 GPa in the post-wolframite phase. They are
presented in Table \ref{tab2} together with the experimental values
measured with Raman spectroscopy by \citet{lacom09} at 26.9 GPa.

\begin{table}
\footnotesize
\small
\renewcommand{\arraystretch}{1.5}
\caption{Calculated and experimental phonon frequencies at $\approx$ 26 GPa and pressure coefficients of the Raman modes for the high-pressure structure of CdWO$_4$.} \centering { \tabcolsep=0.11cm
\begin{tabular}{ccccc}
\hline \hline
       & $\omega_{\text{calc}}$ & $d\omega_{calc}/dP$    & $\omega_{exp}$  & $d\omega_{exp}/dP$    \\  
       & (cm$^{-1}$)     & (cm$^{-1}$/$GPa^{-1}$) & (cm$^{-1}$)     & (cm$^{-1}$/$GPa^{-1}$)  \\ 
symm.  & \multicolumn{2}{c}{26.1 GPa}                    & \multicolumn{2}{c}{26.9 GPa}                     \\  
  \hline
$A_g$  & 57              & 2.90                    & 69              & 1.96                   \\
$B_g$  & 78              & 0.43                    & 88              & 1.94                   \\
$A_g$  & 90              & 0.68                    & 99              & 0.09                   \\
$B_g$  & 122             & 0.04                    & 130             & 0.38                   \\
$B_g$  & 137             & 1.42                    &                 &                        \\
$A_g$  & 150             & -0.49                   & 146             & 1.35                   \\
$A_g$  & 161             & -0.43                   & 155             & 0.97                   \\
$B_g$  & 176             & 1.97                    & 165             & 0.19                   \\
$A_g$  & 178             & 1.20                    & 185             & 1.26                   \\
$B_g$  & 199             & 1.28                    &                 &                        \\
$B_g$  & 218             & 0.91                    & 209             & 1.26                   \\
$A_g$  & 230             & -0.77                   & 243             & -0.06                  \\
$A_g$  & 270             & 2.16                    &                 &                        \\
$B_g$  & 279             & 0.26                    & 279             & 2.53                   \\
$A_g$  & 304             & 2.88                    & 290             & 0.99                   \\
$B_g$  & 306             & 0.19                    & 315             & 3.00                   \\
$A_g$  & 354             & 1.25                    &                 &                        \\
$B_g$  & 380             & 1.46                    & 378             & 1.65                   \\
$A_g$  & 389             & 2.66                    & 401             & 2.31                   \\
$B_g$  & 411             & 3.61                    &                 &                        \\
$A_g$  & 444             & 1.97                    & 428             & 3.03                   \\
$B_g$  & 451             & 1.56                    & 475             & 2.51                   \\
$A_g$  & 474             & 1.90                    & 486             & 2.72                   \\
$B_g$  & 498             & 1.86                    & 512             & 2.33                   \\
$B_g$  & 551             & 2.69                    &                 &                        \\
$A_g$  & 558             & 2.62                    &                 &                        \\
$B_g$  & 590             & 2.43                    & 571             & 2.62                   \\
$A_g$  & 656             & 1.52                    &                 &                        \\
$A_g$  & 673             & 2.39                    & 673             & -0.82                  \\
$A_g$  & 692             & 2.06                    & 688             & 2.81                   \\
$B_g$  & 707             & 1.71                    & 710             & 1.60                   \\
$B_g$  & 727             & 1.38                    &                 &                        \\
$A_g$  & 749             & 1.50                    & 766             & 2.12                   \\
$B_g$  & 803             & 1.85                    &                 &                        \\
$A_g$  & 836             & 1.43                    & 824             & 2.23                   \\
$A_g$  & 881             & 1.45                    & 864             & 2.04                   \\
\hline \hline
\end{tabular}
}
\label{tab2}
\end{table}

\begin{figure}
\centering
\includegraphics[width=0.40\textwidth]{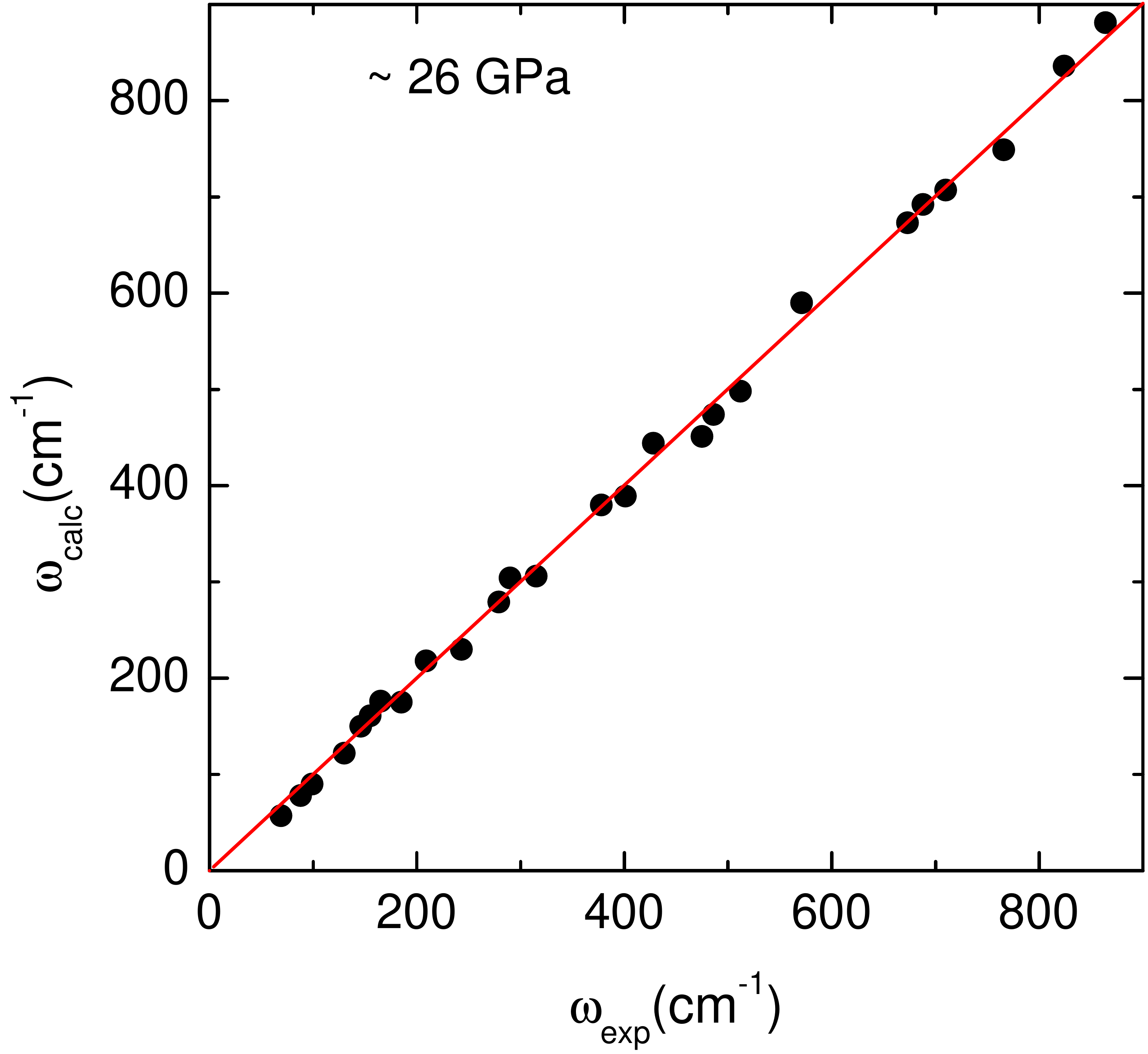}
\caption{\label{fig:fig7} Calculated Raman mode frequencies ($\omega_{calc}$) represented as a function of the experimentally measured values \citep{lacom09} ($\omega_{exp}$) (black points) to show their linear relation with slope 1 (red line).} 
\end{figure}

With the point group $C_{2h}$ and $Z=4$, the post-wolframite structure of CdWO$_4$ (S.G. $P2_1/c$) presents 36
Raman-active modes at zone center $\Gamma = 19 A_g + 17 B_g$. 26 out of those modes had been experimentally
observed \citep{lacom09}. The agreement between the calculated and the experimental
frequencies is excellent as can be directly observed in
Fig. \ref{fig:fig7}. In respect of the pressure coefficients the
calculated and experimental values compare very well for most of the
modes if we consider the overlapping and low intensity of many of the Raman modes
that were measured in the earlier Raman-spectroscopy
study \citep{lacom09}. This confirms that after the phase transition
all the observed Raman modes belong to the single high-pressure post-wolframite phase
and not to two phases as proposed previously \citep{lacom09}.

 \subsection{Electronic Structure}\label{section:bandgap}

\begin{figure}
\centering
\includegraphics[width=0.35\textwidth]{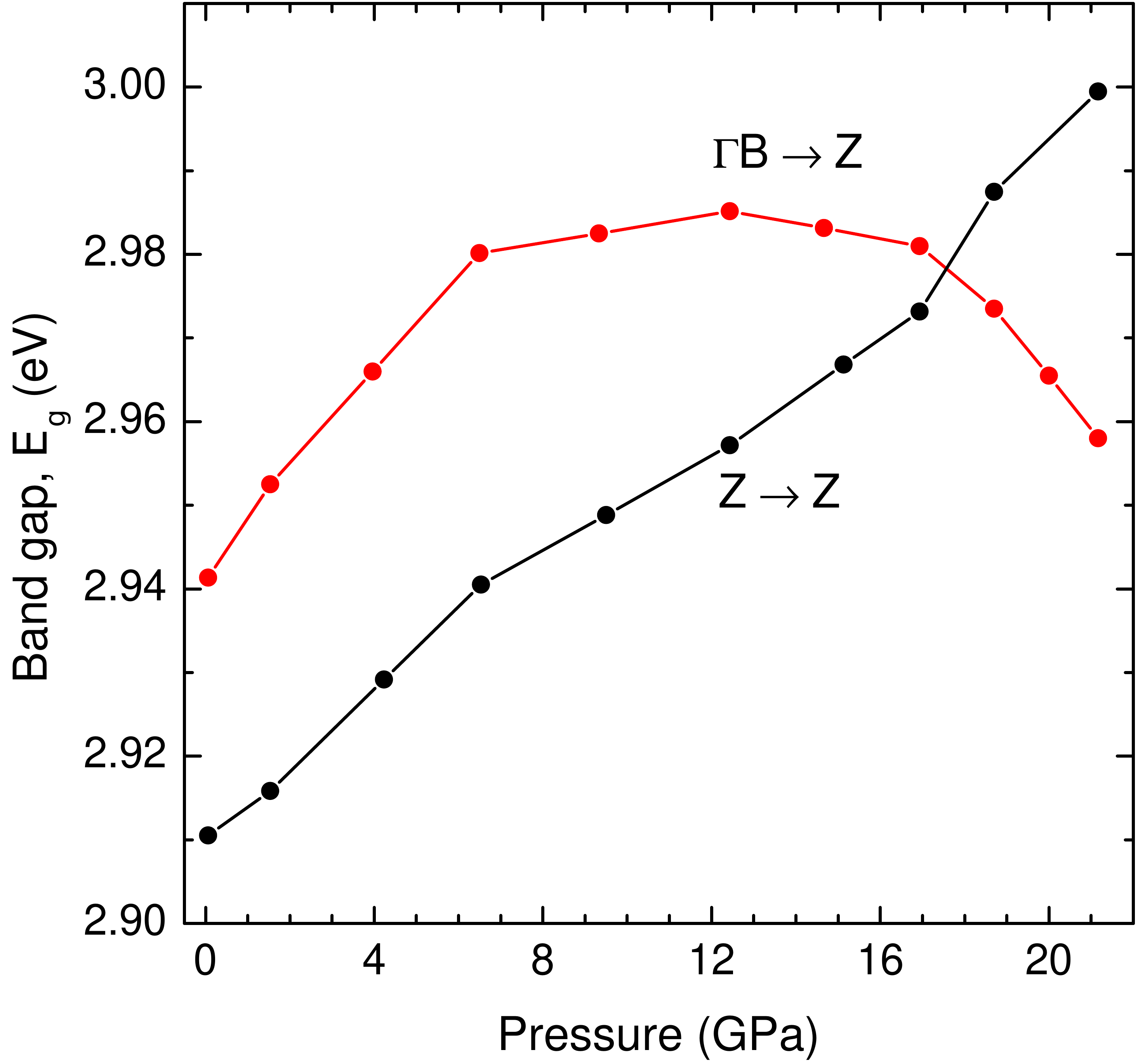}
\caption{\label{fig:fig8} Calculated pressure dependence of the direct ($Z \rightarrow Z$) and indirect ($\Gamma B \rightarrow Z$) band gaps of the low-pressure phase of CdWO$_4$.} 
\end{figure}

Previous calculations had proposed a band crossing in CdWO$_4$ above
10 GPa \citep{ruizf12}. However, the experimental data had been
limited to 9.9 GPa. In this work we have increased the experimental range
to 23 GPa to explore this crossing. As shown in
previous works \citep{abrah00,ruizf12}, the direct band gap of
CdWO$_4$ occurs along the $Z$ point of the Brillouin zone. Calculations underestimate the direct band
gap of the wolframite phase by 1.11 eV, as expected by DFT-GGA
method and predict another indirect band gap $\Gamma B \rightarrow
Z$. Ground states are considered to be accurately determined by the
DFT-GGA. Since both the direct ($Z \rightarrow Z$) and indirect
($\Gamma B \rightarrow Z$) band gaps share the same final state we can
assume that the underestimation (1.11 eV) is the same for both the
direct and indirect transitions. Therefore, at ambient pressure the
indirect band gap of CdWO$_4$ should be at around 4.05 eV, 0.03
eV above the direct band gap and therefore, the indirect transition
cannot be experimentally observed. As pressure is increased
(Fig. \ref{fig:fig8}) the O 2$p$ states (which contribute to the valence
band \citep{ruizf12}) move downwards in energy almost symmetrically at
both $Z$ and $\Gamma B$ points of the Brillouin zone up to 6 GPa. This
results in a parallel energy increase of both direct and indirect
transitions. Above 6 GPa, the energy of the O 2$p$ orbitals continue
decreasing at $Z$ but slow down notably at $\Gamma B$. Above 12 GPa the
energy at $\Gamma B$ increases. This behavior produces an
energy degeneracy of the valence band at both points of the Brillouin
zone at 17 GPa. The result is a band crossing of both direct and
indirect transitions in agreement with the experimental observation at
16.9 GPa (Fig. \ref{fig:fig4}).

\begin{figure*}
\centering
\includegraphics[width=0.60\textwidth]{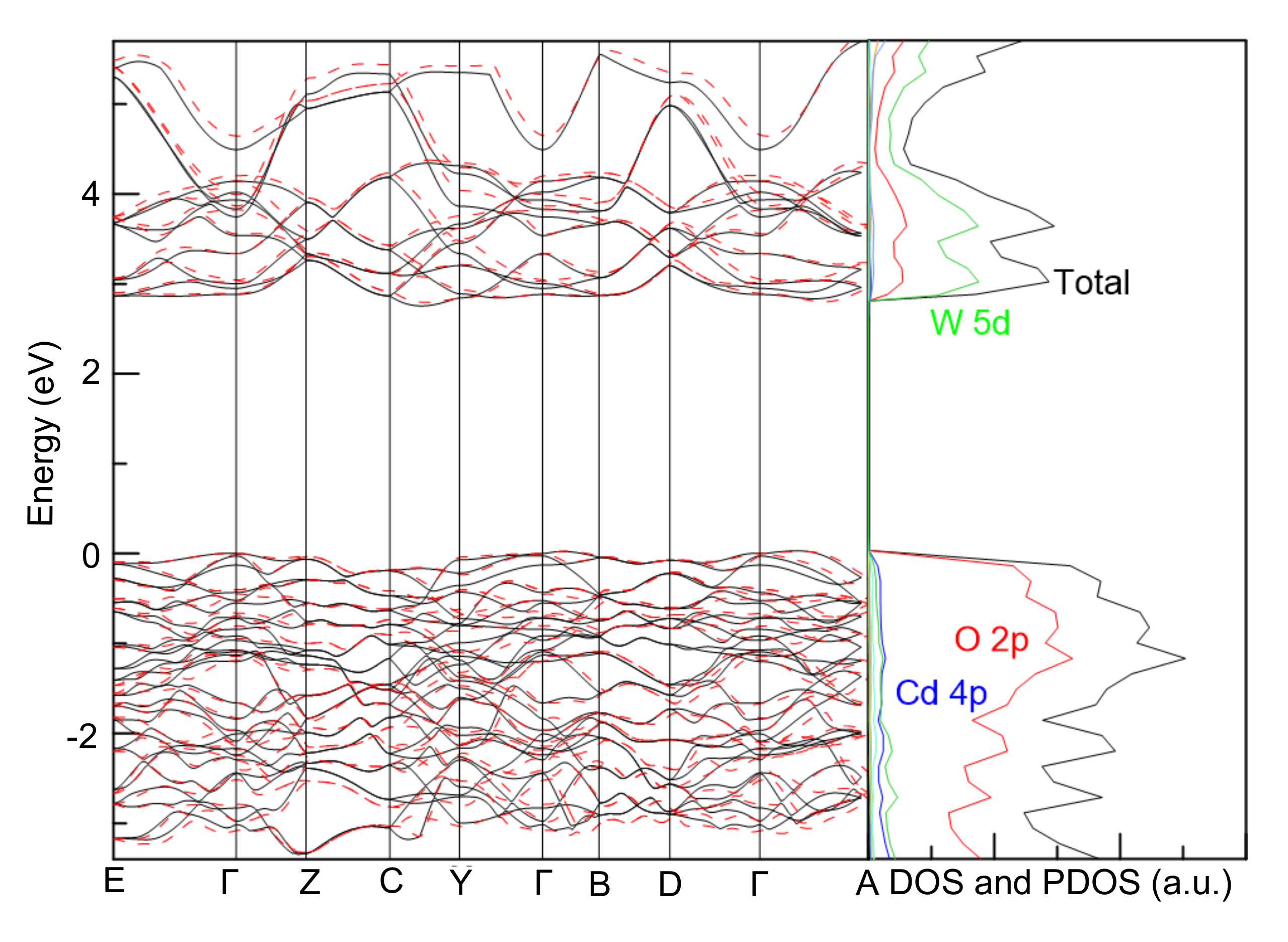}
\caption{\label{fig:fig9} Electronic-band structure of the
  post-wolframite phase of CdWO$_4$ at 18 (black) and 22.8 GPa
  (red). The Fermi level is shown as a dotted line. The right panel shows the corresponding partial and total electron density of states at 18 GPa.}
\end{figure*}

With respect to the high-pressure phase, in Fig. \ref{fig:fig9} we
show the electronic band structure at 18 and 22.8 GPa. Since in the
post-wolframite phase there are two additional non-equivalent
positions for the O atoms and the formula units per unit cell double
from $Z$ = 2 to $Z$ = 4, the number of symmetry directions of the
Brillouin zone increases in two points to $A$ and $E$ giving rise to a more complex band structure with respect to the wolframite phase. Regarding the electronic contribution to the valence and conduction bands they are formed by the O 2$p$ and the W 5$d$ orbitals, respectively, as in the low pressure wolframite phase \citep{ruizf12}.

In spite of the low dispersion of the bands, as a result of the $d$
electrons, a close inspection to the calculated electronic band
structure of the post-wolframite indicates that the maximum of the
valence band is located along $\Gamma A$ and the minimum of the
valence band is found along the $CY$ direction of the Brillouin
zone. This would result into an indirect band gap which would widen with
a pressure coefficient of 7.7 meV/GPa as the result of
the pressure-induced increase in energy of the conduction band. This
value contrasts with the experimental results that indicate a direct
band gap which has a negative pressure coefficient. In fact, the only point of the conduction band that moves downwards in energy with pressure is the point at the $\Gamma A$ direction of the Brillouin zone. When a
direct and an indirect band gap are very close in energy, usually the
direct one is favored. Since the direct transition in $\Gamma A$ is only 0.1 eV higher in energy than the
indirect $\Gamma A \rightarrow CY$, decreases in energy with pressure, and is allowed, we conclude that this direct band gap is the only one observed
experimentally. The calculations underestimate the experimental band gap in $\sim$1 eV similarly to the understimation of the band gap of the low-pressure phase.

\section{CONCLUSIONS}

The high-pressure post-wolframite structure of CdWO$_4$ has been solved at 22 GPa using single-crystal x-ray diffraction. It is a single phase with monoclinic structure and space group $P2_1/c$. Our result contradicts the previous proposal of two coexisting structures \citep{lacom09}. Density-Functional-Theory based calculations support our structural solution which is more stable than the wolframite-type one above 18 GPa and provides a reliable assignment to the Raman modes observed by \citet{lacom09} for the high-pressure phase. Moreover, optical absorption experiments have been extended to 23 GPa confirming previous results \citep{ruizf12} up to 10 GPa. At 16.9 GPa a band crossing of an indirect band gap previously proposed by electronic band structure calculations \citep{ruizf12} within the low-pressure wolframite phase was observed. With regards to the high-pressure phase our optical absorption calculations show a band gap collapse of 0.7 eV as the result of the phase transition to a direct band gap of 3.55 eV at 20 GPa. The experimental observation of a strong redshift (-50 meV) of this high-pressure phase band gap and its direct nature indicate that the calculated indirect $\Gamma A \rightarrow CY$ band gap in post-wolframite is not experimentally observed and instead a direct band gap at $\Gamma A$ appears.

\section*{Acknowledgments}
J.R.-F. thanks the Juan de la Cierva Program (IJCI-2014-20513) of the
Spanish MINECO. A.F. acknowledges financial support from the DFG,
Germany, within priority program SPP1236 (Project
FR-2491/2-1). W.M. acknowledges the BMBF, Germany (Projects 05K10RFA
and 05K13RF1). This paper was partially supported by the Spanish Ministerio de Economıa y Competitividad (MINECO) under grants MAT2013-46649-C04-01/03-P, MAT2016-75586-C4-1/3-P, and No.MAT2015-71070-REDC (MALTA Consolider). Part of this research were carried out at the light
source PETRA III at DESY, a member of the Helmholtz Association (HGF). Portions of this work were performed at HPCAT (Sector 16), Advanced Photon Source (APS), Argonne National Laboratory. HPCAT operations are supported by DOENNSA under Award No. DE-NA0001974 and DOE-BES under Award No. DE-FG02-99ER45775, with partial instrumentation funding by NSF. APS is supported by DOE-BES, under Contract No. DE-AC02-06CH11357.

%\bibliography{prb}

\begin{thebibliography}{38}
\expandafter\ifx\csname natexlab\endcsname\relax\def\natexlab#1{#1}\fi
\expandafter\ifx\csname bibnamefont\endcsname\relax
  \def\bibnamefont#1{#1}\fi
\expandafter\ifx\csname bibfnamefont\endcsname\relax
  \def\bibfnamefont#1{#1}\fi
\expandafter\ifx\csname citenamefont\endcsname\relax
  \def\citenamefont#1{#1}\fi
\expandafter\ifx\csname url\endcsname\relax
  \def\url#1{\texttt{#1}}\fi
\expandafter\ifx\csname urlprefix\endcsname\relax\def\urlprefix{URL }\fi
\providecommand{\bibinfo}[2]{#2}
\providecommand{\eprint}[2][]{\url{#2}}

\bibitem[{\citenamefont{Rathee et~al.}(2006)\citenamefont{Rathee, Tu, Monajemi,
  Rickey, and Fallone}}]{rathe06}
\bibinfo{author}{\bibfnamefont{S.}~\bibnamefont{Rathee}},
  \bibinfo{author}{\bibfnamefont{D.}~\bibnamefont{Tu}},
  \bibinfo{author}{\bibfnamefont{T.~T.} \bibnamefont{Monajemi}},
  \bibinfo{author}{\bibfnamefont{D.~W.} \bibnamefont{Rickey}},
  \bibnamefont{and} \bibinfo{author}{\bibfnamefont{B.~G.}
  \bibnamefont{Fallone}}, \bibinfo{journal}{Med. Phys.}
  \textbf{\bibinfo{volume}{33}}, \bibinfo{pages}{1078} (\bibinfo{year}{2006}).

\bibitem[{\citenamefont{Mikhailik and Kraus}(2010)}]{mikha10}
\bibinfo{author}{\bibfnamefont{V.~B.} \bibnamefont{Mikhailik}}
  \bibnamefont{and} \bibinfo{author}{\bibfnamefont{H.}~\bibnamefont{Kraus}},
  \bibinfo{journal}{Phys. Stat. Sol. B} \textbf{\bibinfo{volume}{247}},
  \bibinfo{pages}{1583} (\bibinfo{year}{2010}).

\bibitem[{\citenamefont{Silva et~al.}(2012)\citenamefont{Silva, Novais, Silva,
  Schmitberger, Macedo, and Bianchi}}]{silva12}
\bibinfo{author}{\bibfnamefont{M.~M.} \bibnamefont{Silva}},
  \bibinfo{author}{\bibfnamefont{S.~M.~V.} \bibnamefont{Novais}},
  \bibinfo{author}{\bibfnamefont{E.~S.~S.} \bibnamefont{Silva}},
  \bibinfo{author}{\bibfnamefont{T.}~\bibnamefont{Schmitberger}},
  \bibinfo{author}{\bibfnamefont{Z.~S.} \bibnamefont{Macedo}},
  \bibnamefont{and} \bibinfo{author}{\bibfnamefont{R.~F.}
  \bibnamefont{Bianchi}}, \bibinfo{journal}{Mater. Sci. Commun.}
  \textbf{\bibinfo{volume}{136}}, \bibinfo{pages}{317} (\bibinfo{year}{2012}).

\bibitem[{\citenamefont{Burachas et~al.}(1996)\citenamefont{Burachas, Danevich,
  Georgadze, Klapdor-Kleingrothaus, Kobychev, Kropivyansky, Kuts, Muller,
  Muzalevsky, Nikolaiko et~al.}}]{burac96}
\bibinfo{author}{\bibfnamefont{S.~P.} \bibnamefont{Burachas}},
  \bibinfo{author}{\bibfnamefont{F.~A.} \bibnamefont{Danevich}},
  \bibinfo{author}{\bibfnamefont{A.~S.} \bibnamefont{Georgadze}},
  \bibinfo{author}{\bibfnamefont{H.~V.} \bibnamefont{Klapdor-Kleingrothaus}},
  \bibinfo{author}{\bibfnamefont{V.~V.} \bibnamefont{Kobychev}},
  \bibinfo{author}{\bibfnamefont{B.~N.} \bibnamefont{Kropivyansky}},
  \bibinfo{author}{\bibfnamefont{V.~N.} \bibnamefont{Kuts}},
  \bibinfo{author}{\bibfnamefont{A.}~\bibnamefont{Muller}},
  \bibinfo{author}{\bibfnamefont{V.~V.} \bibnamefont{Muzalevsky}},
  \bibinfo{author}{\bibfnamefont{A.~S.} \bibnamefont{Nikolaiko}},
  \bibnamefont{et~al.}, \bibinfo{journal}{Nucl. Instrum. and Methods in Phys.
  Research A} \textbf{\bibinfo{volume}{369}}, \bibinfo{pages}{164}
  (\bibinfo{year}{1996}).

\bibitem[{\citenamefont{Danevich et~al.}(2003)\citenamefont{Danevich,
  Georgadze, Kobychev, Nagorny, Nikolaiko, Ponkratenko, Tretyak, Zdesenko,
  Zdesenko, Bizzeti et~al.}}]{danev03}
\bibinfo{author}{\bibfnamefont{F.~A.} \bibnamefont{Danevich}},
  \bibinfo{author}{\bibfnamefont{A.~S.} \bibnamefont{Georgadze}},
  \bibinfo{author}{\bibfnamefont{V.~V.} \bibnamefont{Kobychev}},
  \bibinfo{author}{\bibfnamefont{S.~S.} \bibnamefont{Nagorny}},
  \bibinfo{author}{\bibfnamefont{A.~S.} \bibnamefont{Nikolaiko}},
  \bibinfo{author}{\bibfnamefont{O.~A.} \bibnamefont{Ponkratenko}},
  \bibinfo{author}{\bibfnamefont{V.~I.} \bibnamefont{Tretyak}},
  \bibinfo{author}{\bibfnamefont{S.~Y.} \bibnamefont{Zdesenko}},
  \bibinfo{author}{\bibfnamefont{Y.~G.} \bibnamefont{Zdesenko}},
  \bibinfo{author}{\bibfnamefont{P.~G.} \bibnamefont{Bizzeti}},
  \bibnamefont{et~al.}, \bibinfo{journal}{Phys. Rev. C}
  \textbf{\bibinfo{volume}{67}}, \bibinfo{pages}{014310}
  (\bibinfo{year}{2003}).

\bibitem[{\citenamefont{Kirm et~al.}(2009)\citenamefont{Kirm, Nagirnyi,
  Feldbach, {de Grazia}, Carr\'e, Merdji, Guizard, Geoffroy, Gaudin, Fedorov
  et~al.}}]{kirmm09}
\bibinfo{author}{\bibfnamefont{M.}~\bibnamefont{Kirm}},
  \bibinfo{author}{\bibfnamefont{V.}~\bibnamefont{Nagirnyi}},
  \bibinfo{author}{\bibfnamefont{E.}~\bibnamefont{Feldbach}},
  \bibinfo{author}{\bibfnamefont{M.}~\bibnamefont{{de Grazia}}},
  \bibinfo{author}{\bibfnamefont{B.}~\bibnamefont{Carr\'e}},
  \bibinfo{author}{\bibfnamefont{M.}~\bibnamefont{Merdji}},
  \bibinfo{author}{\bibfnamefont{S.}~\bibnamefont{Guizard}},
  \bibinfo{author}{\bibfnamefont{G.}~\bibnamefont{Geoffroy}},
  \bibinfo{author}{\bibfnamefont{J.}~\bibnamefont{Gaudin}},
  \bibinfo{author}{\bibfnamefont{N.}~\bibnamefont{Fedorov}},
  \bibnamefont{et~al.}, \bibinfo{journal}{Phys. Rev. B}
  \textbf{\bibinfo{volume}{79}}, \bibinfo{pages}{233103}
  (\bibinfo{year}{2009}).

\bibitem[{\citenamefont{Novosad et~al.}(2012)\citenamefont{Novosad, Kostyk,
  Novosad, Luchenko, and Stryganyuk}}]{novos12}
\bibinfo{author}{\bibfnamefont{S.~S.} \bibnamefont{Novosad}},
  \bibinfo{author}{\bibfnamefont{L.~V.} \bibnamefont{Kostyk}},
  \bibinfo{author}{\bibfnamefont{I.~S.} \bibnamefont{Novosad}},
  \bibinfo{author}{\bibfnamefont{A.~P.} \bibnamefont{Luchenko}},
  \bibnamefont{and} \bibinfo{author}{\bibfnamefont{G.~B.}
  \bibnamefont{Stryganyuk}}, \bibinfo{journal}{Acta Phys. Polonica A}
  \textbf{\bibinfo{volume}{122}}, \bibinfo{pages}{717} (\bibinfo{year}{2012}).

\bibitem[{\citenamefont{Abraham et~al.}(2000)\citenamefont{Abraham, Holzwarth,
  and Williams}}]{abrah00}
\bibinfo{author}{\bibfnamefont{Y.}~\bibnamefont{Abraham}},
  \bibinfo{author}{\bibfnamefont{N.~A.~W.} \bibnamefont{Holzwarth}},
  \bibnamefont{and} \bibinfo{author}{\bibfnamefont{R.~T.}
  \bibnamefont{Williams}}, \bibinfo{journal}{Phys. Rev. B}
  \textbf{\bibinfo{volume}{62}}, \bibinfo{pages}{1733} (\bibinfo{year}{2000}).

\bibitem[{\citenamefont{Fujita et~al.}(2008)\citenamefont{Fujita, Itoh,
  Katagiri, Iri, Kitaura, and Mikhailik}}]{fujit08}
\bibinfo{author}{\bibfnamefont{M.}~\bibnamefont{Fujita}},
  \bibinfo{author}{\bibfnamefont{M.}~\bibnamefont{Itoh}},
  \bibinfo{author}{\bibfnamefont{T.}~\bibnamefont{Katagiri}},
  \bibinfo{author}{\bibfnamefont{D.}~\bibnamefont{Iri}},
  \bibinfo{author}{\bibfnamefont{M.}~\bibnamefont{Kitaura}}, \bibnamefont{and}
  \bibinfo{author}{\bibfnamefont{C.~B.} \bibnamefont{Mikhailik}},
  \bibinfo{journal}{Phys. Rev. B} \textbf{\bibinfo{volume}{77}},
  \bibinfo{pages}{155118} (\bibinfo{year}{2008}).

\bibitem[{\citenamefont{Ruiz-Fuertes et~al.}(2012)\citenamefont{Ruiz-Fuertes,
  L\'opez-Moreno, L\'opez-Solano, Errandonea, Segura, Lacomba-Perales,
  Mu{\~{n}}oz, Radescu, Rodr\'iguez-Hern\'andez, Gospodinov et~al.}}]{ruizf12}
\bibinfo{author}{\bibfnamefont{J.}~\bibnamefont{Ruiz-Fuertes}},
  \bibinfo{author}{\bibfnamefont{S.}~\bibnamefont{L\'opez-Moreno}},
  \bibinfo{author}{\bibfnamefont{J.}~\bibnamefont{L\'opez-Solano}},
  \bibinfo{author}{\bibfnamefont{D.}~\bibnamefont{Errandonea}},
  \bibinfo{author}{\bibfnamefont{A.}~\bibnamefont{Segura}},
  \bibinfo{author}{\bibfnamefont{R.}~\bibnamefont{Lacomba-Perales}},
  \bibinfo{author}{\bibfnamefont{A.}~\bibnamefont{Mu{\~{n}}oz}},
  \bibinfo{author}{\bibfnamefont{S.}~\bibnamefont{Radescu}},
  \bibinfo{author}{\bibfnamefont{P.}~\bibnamefont{Rodr\'iguez-Hern\'andez}},
  \bibinfo{author}{\bibfnamefont{M.}~\bibnamefont{Gospodinov}},
  \bibnamefont{et~al.}, \bibinfo{journal}{Phys. Rev. B}
  \textbf{\bibinfo{volume}{86}}, \bibinfo{pages}{125202}
  (\bibinfo{year}{2012}).

\bibitem[{\citenamefont{Jayaraman et~al.}(1995)\citenamefont{Jayaraman, Wang,
  and Sharma}}]{jayar95}
\bibinfo{author}{\bibfnamefont{A.}~\bibnamefont{Jayaraman}},
  \bibinfo{author}{\bibfnamefont{S.~Y.} \bibnamefont{Wang}}, \bibnamefont{and}
  \bibinfo{author}{\bibfnamefont{S.~K.} \bibnamefont{Sharma}},
  \bibinfo{journal}{Current Science} \textbf{\bibinfo{volume}{69}},
  \bibinfo{pages}{44} (\bibinfo{year}{1995}).

\bibitem[{\citenamefont{Lacomba-Perales
  et~al.}(2009)\citenamefont{Lacomba-Perales, Errandonea, Mart\'inez-Garc\'ia,
  Rodr\'iguez-Hern\'andez, Radescu, M\'ujica, Mu{\~{n}}oz, Chervin, and
  Polian}}]{lacom09}
\bibinfo{author}{\bibfnamefont{R.}~\bibnamefont{Lacomba-Perales}},
  \bibinfo{author}{\bibfnamefont{D.}~\bibnamefont{Errandonea}},
  \bibinfo{author}{\bibfnamefont{D.}~\bibnamefont{Mart\'inez-Garc\'ia}},
  \bibinfo{author}{\bibfnamefont{P.}~\bibnamefont{Rodr\'iguez-Hern\'andez}},
  \bibinfo{author}{\bibfnamefont{S.}~\bibnamefont{Radescu}},
  \bibinfo{author}{\bibfnamefont{A.}~\bibnamefont{M\'ujica}},
  \bibinfo{author}{\bibfnamefont{A.}~\bibnamefont{Mu{\~{n}}oz}},
  \bibinfo{author}{\bibfnamefont{J.~C.} \bibnamefont{Chervin}},
  \bibnamefont{and} \bibinfo{author}{\bibfnamefont{A.}~\bibnamefont{Polian}},
  \bibinfo{journal}{Phys. Rev. B} \textbf{\bibinfo{volume}{79}},
  \bibinfo{pages}{094105} (\bibinfo{year}{2009}).

\bibitem[{\citenamefont{Macavei and Schulz}(1993)}]{macav93}
\bibinfo{author}{\bibfnamefont{J.}~\bibnamefont{Macavei}} \bibnamefont{and}
  \bibinfo{author}{\bibfnamefont{H.}~\bibnamefont{Schulz}},
  \bibinfo{journal}{Z. Kristallogr.} \textbf{\bibinfo{volume}{207}},
  \bibinfo{pages}{193} (\bibinfo{year}{1993}).

\bibitem[{\citenamefont{{Jellison Jr.} et~al.}(2011)\citenamefont{{Jellison
  Jr.}, McGuire, Boatner, Budai, Specht, and Singh}}]{jelli11}
\bibinfo{author}{\bibfnamefont{G.~E.} \bibnamefont{{Jellison Jr.}}},
  \bibinfo{author}{\bibfnamefont{M.~A.} \bibnamefont{McGuire}},
  \bibinfo{author}{\bibfnamefont{L.~A.} \bibnamefont{Boatner}},
  \bibinfo{author}{\bibfnamefont{J.~D.} \bibnamefont{Budai}},
  \bibinfo{author}{\bibfnamefont{E.~D.} \bibnamefont{Specht}},
  \bibnamefont{and} \bibinfo{author}{\bibfnamefont{D.~J.} \bibnamefont{Singh}},
  \bibinfo{journal}{Phys. Rev. B} \textbf{\bibinfo{volume}{84}},
  \bibinfo{pages}{195439} (\bibinfo{year}{2011}).

\bibitem[{\citenamefont{Mao et~al.}(1978)\citenamefont{Mao, Bell, Shaner, and
  Steinberg}}]{maohk78}
\bibinfo{author}{\bibfnamefont{H.~K.} \bibnamefont{Mao}},
  \bibinfo{author}{\bibfnamefont{P.~M.} \bibnamefont{Bell}},
  \bibinfo{author}{\bibfnamefont{J.~W.} \bibnamefont{Shaner}},
  \bibnamefont{and} \bibinfo{author}{\bibfnamefont{D.~J.}
  \bibnamefont{Steinberg}}, \bibinfo{journal}{J. Appl. Phys.}
  \textbf{\bibinfo{volume}{49}}, \bibinfo{pages}{3276} (\bibinfo{year}{1978}).

\bibitem[{\citenamefont{Rothkirch et~al.}(2013)\citenamefont{Rothkirch, Gatta,
  Meyer, Merkel, Merlini, and Liermann}}]{rothk13}
\bibinfo{author}{\bibfnamefont{A.}~\bibnamefont{Rothkirch}},
  \bibinfo{author}{\bibfnamefont{G.~D.} \bibnamefont{Gatta}},
  \bibinfo{author}{\bibfnamefont{M.}~\bibnamefont{Meyer}},
  \bibinfo{author}{\bibfnamefont{S.}~\bibnamefont{Merkel}},
  \bibinfo{author}{\bibfnamefont{M.}~\bibnamefont{Merlini}}, \bibnamefont{and}
  \bibinfo{author}{\bibfnamefont{H.~P.} \bibnamefont{Liermann}},
  \bibinfo{journal}{J. Synchrotron Rad.} \textbf{\bibinfo{volume}{20}},
  \bibinfo{pages}{711} (\bibinfo{year}{2013}).

\bibitem[{\citenamefont{Agilent}(2013)}]{crysal}
\bibinfo{author}{\bibnamefont{Agilent}}, \emph{\bibinfo{title}{Crysalis$^{Pro}$
  software system}}, \bibinfo{howpublished}{version 1.171.36.28, Agilent
  Technologies UK Ltd., Oxford, UK} (\bibinfo{year}{2013}).

\bibitem[{\citenamefont{Sheldrick}(2008)}]{shel08}
\bibinfo{author}{\bibfnamefont{G.~M.} \bibnamefont{Sheldrick}},
  \bibinfo{journal}{Acta Crystallogr. A} \textbf{\bibinfo{volume}{64}},
  \bibinfo{pages}{112} (\bibinfo{year}{2008}).

\bibitem[{\citenamefont{Spek}(2009)}]{spek2009}
\bibinfo{author}{\bibfnamefont{A.~L.} \bibnamefont{Spek}},
  \bibinfo{journal}{Acta Cryst. D} \textbf{\bibinfo{volume}{65}},
  \bibinfo{pages}{148} (\bibinfo{year}{2009}).

\bibitem[{\citenamefont{Farrugia}(1999)}]{farrugia1999}
\bibinfo{author}{\bibfnamefont{L.~J.} \bibnamefont{Farrugia}},
  \bibinfo{journal}{J. Appl. Crystallogr.} \textbf{\bibinfo{volume}{32}},
  \bibinfo{pages}{837} (\bibinfo{year}{1999}).

\bibitem[{\citenamefont{Hohenberg and Kohn}(1996)}]{hohen64}
\bibinfo{author}{\bibfnamefont{P.}~\bibnamefont{Hohenberg}} \bibnamefont{and}
  \bibinfo{author}{\bibfnamefont{W.}~\bibnamefont{Kohn}},
  \bibinfo{journal}{Phys. Rev.} \textbf{\bibinfo{volume}{139}},
  \bibinfo{pages}{864} (\bibinfo{year}{1996}).

\bibitem[{\citenamefont{Kresse and Hafner}(1993)}]{kress93}
\bibinfo{author}{\bibfnamefont{G.}~\bibnamefont{Kresse}} \bibnamefont{and}
  \bibinfo{author}{\bibfnamefont{J.}~\bibnamefont{Hafner}},
  \bibinfo{journal}{Phys. Rev. B} \textbf{\bibinfo{volume}{47}},
  \bibinfo{pages}{558} (\bibinfo{year}{1993}).

\bibitem[{\citenamefont{Kresse and Hafner}(1994)}]{kress94}
\bibinfo{author}{\bibfnamefont{G.}~\bibnamefont{Kresse}} \bibnamefont{and}
  \bibinfo{author}{\bibfnamefont{J.}~\bibnamefont{Hafner}},
  \bibinfo{journal}{Phys. Rev. B} \textbf{\bibinfo{volume}{49}},
  \bibinfo{pages}{14251} (\bibinfo{year}{1994}).

\bibitem[{\citenamefont{Kresse and Furthm\"uller}(1996)}]{kress96}
\bibinfo{author}{\bibfnamefont{G.}~\bibnamefont{Kresse}} \bibnamefont{and}
  \bibinfo{author}{\bibfnamefont{J.}~\bibnamefont{Furthm\"uller}},
  \bibinfo{journal}{Phys. Rev. B} \textbf{\bibinfo{volume}{54}},
  \bibinfo{pages}{11169} (\bibinfo{year}{1996}).

\bibitem[{\citenamefont{Perdew et~al.}(1996)\citenamefont{Perdew, Burke, and
  Ernzerhof}}]{perde96}
\bibinfo{author}{\bibfnamefont{J.}~\bibnamefont{Perdew}},
  \bibinfo{author}{\bibfnamefont{K.}~\bibnamefont{Burke}}, \bibnamefont{and}
  \bibinfo{author}{\bibfnamefont{M.}~\bibnamefont{Ernzerhof}},
  \bibinfo{journal}{Phys. Rev. Lett.} \textbf{\bibinfo{volume}{77}},
  \bibinfo{pages}{3865} (\bibinfo{year}{1996}).

\bibitem[{\citenamefont{Bl\"och}(1994)}]{bloch94}
\bibinfo{author}{\bibfnamefont{P.~E.} \bibnamefont{Bl\"och}},
  \bibinfo{journal}{Phys. Rev. B} \textbf{\bibinfo{volume}{50}},
  \bibinfo{pages}{17953} (\bibinfo{year}{1994}).

\bibitem[{\citenamefont{Kresse and Joubert}(1999)}]{kress99}
\bibinfo{author}{\bibfnamefont{G.}~\bibnamefont{Kresse}} \bibnamefont{and}
  \bibinfo{author}{\bibfnamefont{D.}~\bibnamefont{Joubert}},
  \bibinfo{journal}{Phys. Rev. B} \textbf{\bibinfo{volume}{59}},
  \bibinfo{pages}{1758} (\bibinfo{year}{1999}).

\bibitem[{\citenamefont{Monkhorst and Pack}(1976)}]{monkh76}
\bibinfo{author}{\bibfnamefont{H.~J.} \bibnamefont{Monkhorst}}
  \bibnamefont{and} \bibinfo{author}{\bibfnamefont{J.~D.} \bibnamefont{Pack}},
  \bibinfo{journal}{Phys. Rev. B} \textbf{\bibinfo{volume}{13}},
  \bibinfo{pages}{5188} (\bibinfo{year}{1976}).

\bibitem[{\citenamefont{Ruiz-Fuertes et~al.}(2010)\citenamefont{Ruiz-Fuertes,
  L\'opez-Moreno, Errandonea, Pellicer-Porres, Lacomba-Perales, Segura,
  Rodr\'iguez-Hern\'andez, Mu{\~{n}}oz, Romero, and Gonz\'alez}}]{ruizf10}
\bibinfo{author}{\bibfnamefont{J.}~\bibnamefont{Ruiz-Fuertes}},
  \bibinfo{author}{\bibfnamefont{S.}~\bibnamefont{L\'opez-Moreno}},
  \bibinfo{author}{\bibfnamefont{D.}~\bibnamefont{Errandonea}},
  \bibinfo{author}{\bibfnamefont{J.}~\bibnamefont{Pellicer-Porres}},
  \bibinfo{author}{\bibfnamefont{R.}~\bibnamefont{Lacomba-Perales}},
  \bibinfo{author}{\bibfnamefont{A.}~\bibnamefont{Segura}},
  \bibinfo{author}{\bibfnamefont{P.}~\bibnamefont{Rodr\'iguez-Hern\'andez}},
  \bibinfo{author}{\bibfnamefont{A.}~\bibnamefont{Mu{\~{n}}oz}},
  \bibinfo{author}{\bibfnamefont{A.~H.} \bibnamefont{Romero}},
  \bibnamefont{and}
  \bibinfo{author}{\bibfnamefont{J.}~\bibnamefont{Gonz\'alez}},
  \bibinfo{journal}{J. Appl. Phys.} \textbf{\bibinfo{volume}{107}},
  \bibinfo{pages}{083506} (\bibinfo{year}{2010}).

\bibitem[{\citenamefont{Urbach}(1953)}]{urbac53}
\bibinfo{author}{\bibfnamefont{F.}~\bibnamefont{Urbach}},
  \bibinfo{journal}{Phys. Rev.} \textbf{\bibinfo{volume}{92}},
  \bibinfo{pages}{1324} (\bibinfo{year}{1953}).

\bibitem[{\citenamefont{Klotz et~al.}(2009)\citenamefont{Klotz, Chervin,
  Munsch, and Marchand}}]{klotz09}
\bibinfo{author}{\bibfnamefont{S.}~\bibnamefont{Klotz}},
  \bibinfo{author}{\bibfnamefont{J.-C.} \bibnamefont{Chervin}},
  \bibinfo{author}{\bibfnamefont{P.}~\bibnamefont{Munsch}}, \bibnamefont{and}
  \bibinfo{author}{\bibfnamefont{G.~L.} \bibnamefont{Marchand}},
  \bibinfo{journal}{J. Phys. D: Appl. Phys.} \textbf{\bibinfo{volume}{42}},
  \bibinfo{pages}{075413} (\bibinfo{year}{2009}).

\bibitem[{\citenamefont{Ruiz-Fuertes et~al.}(2008)\citenamefont{Ruiz-Fuertes,
  Errandonea, Manj\'on, Mart\'inez-Garc\'ia, Segura, Ursak, and
  Tiginyanu}}]{ruizf08}
\bibinfo{author}{\bibfnamefont{J.}~\bibnamefont{Ruiz-Fuertes}},
  \bibinfo{author}{\bibfnamefont{D.}~\bibnamefont{Errandonea}},
  \bibinfo{author}{\bibfnamefont{F.~J.} \bibnamefont{Manj\'on}},
  \bibinfo{author}{\bibfnamefont{D.}~\bibnamefont{Mart\'inez-Garc\'ia}},
  \bibinfo{author}{\bibfnamefont{A.}~\bibnamefont{Segura}},
  \bibinfo{author}{\bibfnamefont{V.~V.} \bibnamefont{Ursak}}, \bibnamefont{and}
  \bibinfo{author}{\bibfnamefont{I.~M.} \bibnamefont{Tiginyanu}},
  \bibinfo{journal}{J. Appl. Phys.} \textbf{\bibinfo{volume}{103}},
  \bibinfo{pages}{063710} (\bibinfo{year}{2008}).

\bibitem[{\citenamefont{Errandonea et~al.}(2006)\citenamefont{Errandonea,
  Mart\'inez-Garc\'ia, Lacomba-Perales, Ruiz-Fuertes, and Segura}}]{erran06}
\bibinfo{author}{\bibfnamefont{D.}~\bibnamefont{Errandonea}},
  \bibinfo{author}{\bibfnamefont{D.}~\bibnamefont{Mart\'inez-Garc\'ia}},
  \bibinfo{author}{\bibfnamefont{R.}~\bibnamefont{Lacomba-Perales}},
  \bibinfo{author}{\bibfnamefont{J.}~\bibnamefont{Ruiz-Fuertes}},
  \bibnamefont{and} \bibinfo{author}{\bibfnamefont{A.}~\bibnamefont{Segura}},
  \bibinfo{journal}{Appl. Phys. Lett.} \textbf{\bibinfo{volume}{89}},
  \bibinfo{pages}{091913} (\bibinfo{year}{2006}).

\bibitem[{\citenamefont{Lacomba-Perales
  et~al.}(2011)\citenamefont{Lacomba-Perales, Errandonea, Segura, Ruiz-Fuertes,
  Rodr\'iguez-Hern\'andez, Radescu, L\'opez-Moreno, M\'ujica, and
  Mu{\~{n}}oz}}]{lacom11}
\bibinfo{author}{\bibfnamefont{R.}~\bibnamefont{Lacomba-Perales}},
  \bibinfo{author}{\bibfnamefont{D.}~\bibnamefont{Errandonea}},
  \bibinfo{author}{\bibfnamefont{A.}~\bibnamefont{Segura}},
  \bibinfo{author}{\bibfnamefont{J.}~\bibnamefont{Ruiz-Fuertes}},
  \bibinfo{author}{\bibfnamefont{P.}~\bibnamefont{Rodr\'iguez-Hern\'andez}},
  \bibinfo{author}{\bibfnamefont{S.}~\bibnamefont{Radescu}},
  \bibinfo{author}{\bibfnamefont{S.}~\bibnamefont{L\'opez-Moreno}},
  \bibinfo{author}{\bibfnamefont{A.}~\bibnamefont{M\'ujica}}, \bibnamefont{and}
  \bibinfo{author}{\bibfnamefont{A.}~\bibnamefont{Mu{\~{n}}oz}},
  \bibinfo{journal}{J. Appl. Phys.} \textbf{\bibinfo{volume}{110}},
  \bibinfo{pages}{043703} (\bibinfo{year}{2011}).

\bibitem[{\citenamefont{Panchal et~al.}(2011)\citenamefont{Panchal, Errandonea,
  Segura, Rodr\'iguez-Hern\'andez, Mu{\~{n}}oz, L\'opez-Moreno, and
  Bettinelli}}]{panch11}
\bibinfo{author}{\bibfnamefont{V.}~\bibnamefont{Panchal}},
  \bibinfo{author}{\bibfnamefont{D.}~\bibnamefont{Errandonea}},
  \bibinfo{author}{\bibfnamefont{A.}~\bibnamefont{Segura}},
  \bibinfo{author}{\bibfnamefont{P.}~\bibnamefont{Rodr\'iguez-Hern\'andez}},
  \bibinfo{author}{\bibfnamefont{A.}~\bibnamefont{Mu{\~{n}}oz}},
  \bibinfo{author}{\bibfnamefont{S.}~\bibnamefont{L\'opez-Moreno}},
  \bibnamefont{and}
  \bibinfo{author}{\bibfnamefont{M.}~\bibnamefont{Bettinelli}},
  \bibinfo{journal}{J. Appl. Phys.} \textbf{\bibinfo{volume}{110}},
  \bibinfo{pages}{043723} (\bibinfo{year}{2011}).

\bibitem[{\citenamefont{Errandonea et~al.}(2016)\citenamefont{Errandonea,
  Popescu, Garg, Botella, Mart\'inez-Garc\'ia, Pellicer-Porres,
  Rodr\'iguez-Hern\'andez, Mu{\~{n}}oz, Cuenca-Gotor, and Sans}}]{erran16}
\bibinfo{author}{\bibfnamefont{D.}~\bibnamefont{Errandonea}},
  \bibinfo{author}{\bibfnamefont{C.}~\bibnamefont{Popescu}},
  \bibinfo{author}{\bibfnamefont{A.~B.} \bibnamefont{Garg}},
  \bibinfo{author}{\bibfnamefont{P.}~\bibnamefont{Botella}},
  \bibinfo{author}{\bibfnamefont{D.}~\bibnamefont{Mart\'inez-Garc\'ia}},
  \bibinfo{author}{\bibfnamefont{J.}~\bibnamefont{Pellicer-Porres}},
  \bibinfo{author}{\bibfnamefont{P.}~\bibnamefont{Rodr\'iguez-Hern\'andez}},
  \bibinfo{author}{\bibfnamefont{A.}~\bibnamefont{Mu{\~{n}}oz}},
  \bibinfo{author}{\bibfnamefont{V.}~\bibnamefont{Cuenca-Gotor}},
  \bibnamefont{and} \bibinfo{author}{\bibfnamefont{J.~A.} \bibnamefont{Sans}},
  \bibinfo{journal}{Phys. Rev. B} \textbf{\bibinfo{volume}{93}},
  \bibinfo{pages}{035204} (\bibinfo{year}{2016}).

\bibitem[{sup()}]{support}
\bibinfo{note}{See Supplemental Material at [URL will be inserted by publisher]
  in order to see the powder diffraction patterns of CdWO$_4$ at 1.3 and 13.2
  GPa and a detailed table with the SXRD data collection parameters and
  refinement results including the equivalent or isotropic displacement
  parameters at 15 GPa for the low-pressure wolframite-type phase and at 22 GPa
  for the high-pressure post-wolframite phase.}

\bibitem[{\citenamefont{Ruiz-Fuertes et~al.}(2015)\citenamefont{Ruiz-Fuertes,
  Friedrich, Gomis, Errandonea, Morgenroth, Sans, and
  Santamar\'ia-P\'erez}}]{ruizf15}
\bibinfo{author}{\bibfnamefont{J.}~\bibnamefont{Ruiz-Fuertes}},
  \bibinfo{author}{\bibfnamefont{A.}~\bibnamefont{Friedrich}},
  \bibinfo{author}{\bibfnamefont{O.}~\bibnamefont{Gomis}},
  \bibinfo{author}{\bibfnamefont{D.}~\bibnamefont{Errandonea}},
  \bibinfo{author}{\bibfnamefont{W.}~\bibnamefont{Morgenroth}},
  \bibinfo{author}{\bibfnamefont{J.~A.} \bibnamefont{Sans}}, \bibnamefont{and}
  \bibinfo{author}{\bibfnamefont{D.}~\bibnamefont{Santamar\'ia-P\'erez}},
  \bibinfo{journal}{Phys. Rev. B} \textbf{\bibinfo{volume}{91}},
  \bibinfo{pages}{104109} (\bibinfo{year}{2015}).

\end{thebibliography}

\end{document}